\documentclass[preprint,12pt,numbers,sort&compress]{elsarticle}

\usepackage{amssymb}
\usepackage{amsmath}
\usepackage{mathrsfs}
\usepackage{graphicx}
\usepackage{hyperref}
\hypersetup{hidelinks, colorlinks=true, allcolors=blue,}
\usepackage{soul}
\usepackage{xcolor}
\soulregister\cite7
\soulregister\ref7

\journal{Elsevier}
\begin{document}

\begin{frontmatter}

\title{An improved fully one-sided diffuse-interface immersed boundary method with target-value reconstruction for compressible flows}

\author{Qian Mao\corref{cor1}}
\cortext[cor1]{Corresponding author: qian.mao@univ-amu.fr}
\author{Song Zhao}
\author{Pierre Boivin}
\author{Julien Favier}
\affiliation{organization={Aix-Marseille Univ, CNRS, Centrale Med, M2P2}, city={Marseille}, country={France}}

\begin{abstract}

Although one-sided spreading has been shown to improve the near-wall accuracy of diffuse-interface immersed boundary methods (DIBMs), the effect of its asymmetric kernel support on the effective boundary location remains insufficiently understood. In this work, a detailed analysis of the one-sided spreading operator reveals an inward displacement of the effective boundary relative to the geometric boundary. To compensate for this displacement, a target-value reconstruction strategy is developed to ensure consistency between the values imposed at the effective boundary and the prescribed conditions at the geometric boundary. The strategy is incorporated into the fully one-sided diffuse-interface immersed boundary method (FODIBM) and applies to both Dirichlet and Neumann boundary conditions. Although confined to the target-value evaluation step, the modification substantially improves boundary-condition enforcement with negligible additional computational cost. Coupled with a hybrid lattice Boltzmann solver, the improved method consistently reduces $\mathrm{L}_2$ and $\mathrm{L}_{\infty}$ error norms across different grid resolutions while retaining approximately second-order grid convergence. The no-slip and isothermal boundary-condition errors are reduced by $77\%$ and $85\%$, respectively. Simulations involving various two- and three-dimensional geometries further show improved predictions relative to both the conventional DIBM and the original FODIBM. The results agree well with body-fitted reference solutions and experimental data, demonstrating accurate and computationally efficient simulations of compressible flows around complex geometries.

\end{abstract}

\begin{keyword}
Diffuse-interface immersed boundary method, Lattice Boltzmann method, Compressible flows, Complex geometries
\end{keyword}

\end{frontmatter}

\section{Introduction}
\label{Intro}

Compressible flow around complex geometries lies at the core of a wide range of engineering applications, such as high-speed vehicles and spacecraft reentry. Computational fluid dynamics (CFD) plays a pivotal role in investigating these problems. Conventional methods predominantly rely on body-fitted meshes. While these methods allow accurate representation of wall boundaries, they make it difficult to generate high-quality grids for complex geometries, even with substantial advances in mesh generation techniques over recent decades. The immersed boundary method (IBM) has emerged as an attractive alternative due to its simplicity and effectiveness in handling boundary conditions. IBMs represent the effects of geometric boundaries as forcing source terms in the governing equations. This strategy avoids the need for body-fitted meshes and enables simulations on simple Cartesian grids. As a result, mesh generation is substantially simplified and computational flexibility is enhanced.

IBMs can generally be classified into sharp-interface and diffuse-interface approaches. Sharp-interface IBMs impose forcing terms directly on discrete fluid nodes in the vicinity of the boundary. They require elaborate treatments near the boundary, such as cell classification, trimming, or reshaping \cite{mittal2008versatile,schneiders2013accurate,jost2021direct}. These strategies enable accurate enforcement of boundary conditions, but they become complicated for complex moving geometries and may introduce numerical instability. In contrast, diffuse-interface IBMs rely on an interpolation and spreading strategy using a smoothed regularized delta function \cite{peskin2002immersed,mao2026fluid,mao2024hydrodynamic}. This approach enforces boundary conditions in a simple and efficient manner. As a result, diffuse-interface IBMs show strong potential for handling complex geometries. The original diffuse-interface IBM was proposed by Peskin \cite{peskin1972flow} to simulate the interaction between heart valves and blood flow. The boundary is represented by a set of Lagrangian points, while the flow is solved on a Cartesian mesh, referred to as Eulerian points. In this approach, the internal elastic force that balances the fluid force serves as the forcing term. Consequently, it is not suitable for rigid boundaries. To overcome this limitation, Goldstein et al. \cite{goldstein1993modeling} proposed a feedback or penalty IBM. A system of virtual springs and dampers is used to evaluate the forcing term and enforce the no-slip boundary condition. This method requires tuning of damping parameters for different problems and imposes a restriction on the time step to reduce oscillations. To eliminate such free parameters, Fadlun et al. \cite{fadlun2000combined} introduced a direct forcing IBM. The forcing term is computed from the difference between the desired boundary velocity and the flow velocity at the boundary points (Eulerian points) and applied directly. Subsequently, Uhlmann \cite{uhlmann2005immersed} combined this approach with the regularized delta function. The forcing term is computed at the Lagrangian points and then smoothly distributed to the surrounding Eulerian points. This improvement effectively suppresses the numerical oscillations. Due to the non-reciprocity between the interpolation and spreading operators, an explicit one-step forcing strategy cannot accurately enforce the velocity boundary condition. Several approaches have been proposed to improve accuracy. These include the multi-direct forcing IBM, which applies forcing multiple times within a single time step \cite{luo2007full,wang2008combined,zhang2020relaxed}; the implicit IBM, which solves a linear system to better satisfy the reciprocity condition \cite{wu2009implicit,wu2010improved}; and methods based on Lagrangian weights, which approximately restore reciprocity and explicitly correct the forcing term \cite{gsell2019explicit,gsell2021direct,cheylan2023analysis}.

Despite the above advances, most IBMs have been developed for incompressible flows and consider only Dirichlet boundary conditions for velocity \cite{mao2021hydrodynamic,mao2023snap}. To extend IBMs to compressible flows and improve their generality, thermal boundary conditions must also be incorporated, such as isothermal (Dirichlet) and adiabatic (Neumann) conditions. Several efforts have been made in this direction, including extensions of the feedback IBM \cite{wang2017immersed,menez2023assessment}, the direct forcing IBM \cite{riahi2018pressure,midani2025immersed,lerogeron2025numerical}, and the implicit IBM \cite{qiu2016boundary,sun2020diffuse}. However, several issues arise due to the inherent diffusion effect of diffuse-interface IBMs. This effect refers to the smearing of the boundary caused by the smooth distribution of the forcing term over Eulerian points near the interface. As a consequence, a spurious pressure drop appears near the boundary, resulting in an underestimation of the surface pressure coefficient \cite{menez2023assessment,qiu2016boundary}. In addition, the enforcement of adiabatic boundary conditions is often inaccurate \cite{lerogeron2025numerical}. Various strategies have been proposed to mitigate this diffusion effect. One approach introduces an inward retraction of the Lagrangian points \cite{breugem2012second,luo2019improved}. The optimal retraction distance depends on the Reynolds number and grid resolution. While this method performs well for incompressible particle-laden flows, its effectiveness for complex geometries remains to be fully validated. Another approach is the half-distribution forcing strategy \cite{ji2012novel,peng2020force}. This method employs a two-sided interpolation and one-sided spreading scheme, in which the forcing term is applied only inside the boundary, thereby effectively reducing the diffusion effect. Wu et al. \cite{wu2025one} extended this strategy to compressible flows by combining it with an implicit IBM. This formulation eliminates the spurious pressure drop and allows for accurate enforcement of boundary conditions. More recently, Mao et al. \cite{mao2026explicit} developed an explicit fully one-sided diffuse-interface IBM, in which both interpolation and spreading are performed exclusively on the interior side of the boundary. This approach offers significantly improved stability for compressible flows compared to the explicit half-distribution forcing strategy, and avoids solving the linear system required by the implicit IBM. It also demonstrates good performance across a wide range of complex flow configurations.

Although one-sided spreading schemes developed in previous studies substantially improve the near-wall accuracy of diffuse-interface IBMs, the effect of their asymmetric kernel support on the effective boundary representation remains insufficiently understood. In this study, a quantitative analysis reveals that the asymmetric support of the one-sided spreading operator causes an inward displacement of the effective boundary relative to the geometric boundary. Based on this analysis, a target-value reconstruction strategy is developed within the fully one-sided diffuse-interface IBM to improve the enforcement of boundary conditions without altering the original interpolation and spreading procedures. Although confined to the target-value evaluation step, the proposed reconstruction substantially improves boundary-condition enforcement at negligible additional computational cost. The resulting method is evaluated through numerical assessments of convergence behavior, boundary-condition accuracy, and computational cost, followed by a series of two- and three-dimensional compressible-flow cases involving complex geometries, with particular emphasis on near-wall predictions and shock locations.

The remainder of this paper is organized as follows. Section \ref{Num} presents the governing equations, the compressible flow solver, and the improved fully one-sided diffuse-interface IBM with target-value reconstruction. Section \ref{res} presents and discusses the accuracy and validation results. Finally, conclusions are drawn in Section \ref{conclu}.

\section{Numerical methods}
\label{Num}

The flow under consideration is described by the compressible Navier-Stokes equations, which arise from the conservation of mass, momentum, and energy. To solve these equations, a hybrid lattice Boltzmann method (LBM) framework combined with the total energy equation is adopted \cite{farag2021unified,wissocq2022restoring}. Boundary conditions of both Dirichlet and Neumann types are enforced via a fully one-sided diffuse-interface immersed boundary method incorporating a target-value reconstruction strategy. A detailed description of the methodology is provided below.

\subsection{Governing equations}
\label{sub_govequ}

The compressible Navier-Stokes (N-S) equations are expressed as:

\begin{equation}
\frac{\partial\mathbf{Q}}{\partial t}+\nabla\cdot\mathbf{F}=\mathbf{S},
\end{equation}

\begin{equation}
\left.\mathbf{Q}=\left\{\begin{array}{c}\rho\\\rho\boldsymbol{u}\\\rho E\end{array}\right.\right\},\quad\mathbf{F}=\begin{Bmatrix}\rho\boldsymbol{u}\\\rho\boldsymbol{u}\otimes\boldsymbol{u}+p\boldsymbol{I}+\boldsymbol{\tau}\\(\rho E+p)\boldsymbol{u}+\boldsymbol{u}\cdot\boldsymbol{\tau}+\boldsymbol{q}\end{Bmatrix},\quad\mathbf{S}=\begin{Bmatrix}0\\\boldsymbol{f}_{\boldsymbol{u}}\\f_{E}\end{Bmatrix},
\end{equation}

\begin{equation}
\boldsymbol{\tau}=-\mu\left[\left(\nabla\boldsymbol{u}+(\nabla\boldsymbol{u})^\mathrm{T}\right)-\frac{2}{3}(\nabla\cdot\boldsymbol{u})\boldsymbol{I}\right],
\end{equation}

\begin{equation}
\boldsymbol{q}=-\lambda\nabla T,
\end{equation}
where $\rho$, $\boldsymbol{u}$, $E$, $p$, $\boldsymbol{\tau}$, $T$, $\mu$, $\lambda$, $\boldsymbol{f}_{\boldsymbol{u}}$ and $f_E$ denote the flow density, the velocity, the total energy, the pressure, the viscous stress tensor, the temperature, the dynamic viscosity, the thermal conductivity, the momentum forcing term and the energy forcing term, respectively. The pressure, density, and temperature are related through the ideal gas law, $p = \rho R T$, where $R$ denotes the specific gas constant. The total energy is expressed as the sum of kinetic and internal contributions, $E = |\boldsymbol{u}|^2/2 + C_v T$, with $C_v$ representing the specific heat capacity at constant volume.

\subsection{Hybrid LBM with the total energy equation}
\label{sub_LB}

In the LBM framework, the fluid is described by the particle distribution function $f(\boldsymbol{x},\boldsymbol{\zeta},t)$, which denotes the probability density of fluid particles with velocity $\boldsymbol{\zeta}$ at position $\boldsymbol{x}$ and time $t$. The evolution of $f(\boldsymbol{x},\boldsymbol{\zeta},t)$ is governed by the Boltzmann equation:

\begin{equation}
    \frac{\partial f}{\partial t} + \boldsymbol{\zeta} \cdot  \nabla f= \Gamma (f),
\label{eq_bol}
\end{equation}
where $\Gamma$ is the collision operator. The velocity space is discretized by a set of velocity vectors \{$\boldsymbol c_i$, $i=0$, \ldots, $Q-1$\}, where $Q$ denotes the total number of discrete velocities. The $D3Q19$ scheme is employed, and the corresponding discrete velocities are defined in Eq. (\ref{eq_cid3q19}) of \ref{app_d3q19}. The lattice Boltzmann equation discretized in velocity space, physical space and time is given by \cite{farag2021unified}

\begin{equation}
\begin{aligned}
    \bar{f}_i(\boldsymbol{x},t+\Delta t) &= f^{eq}_i(\boldsymbol{x}-\boldsymbol{c}_{i}\Delta t,t) + (1-\frac{\Delta t}{\bar{\tau}_{\scriptscriptstyle \mathrm{LB}}})\bar{f}^{neq}_i(\boldsymbol{x}-\boldsymbol{c}_{i}\Delta t,t) \\
    &+ \frac{\Delta t}{2}F_i(\boldsymbol{x}-\boldsymbol{c}_{i}\Delta t,t),
\end{aligned}
\label{eq_LB}
\end{equation} 
where $\bar{f}_i$ is defined as $\bar{f}_i=f_i -[(\Gamma_i+F_i)\Delta t]/2$. $\bar{\tau}_{\scriptscriptstyle \mathrm{LB}}$ is defined as $\bar{\tau}_{\scriptscriptstyle \mathrm{LB}}=\tau_{\scriptscriptstyle \mathrm{LB}} +\Delta t/2$, where $\tau_{\scriptscriptstyle \mathrm{LB}}=\nu / c_s^2$ is the relaxation time and $\nu$ denotes the kinematic viscosity. $f^{eq}_i$ and $\bar{f}^{neq}_i$ denote the equilibrium and non-equilibrium distribution functions, as defined in Eqs. (\ref{eq_feq}) and (\ref{eq_fneq}) of Appendix \ref{app_d3q19}. The external body force term $F_i$ is defined as

\begin{equation}
    F_i = w_i\left[ \frac{\mathcal{H}_{i,\alpha}^{(1)}}{c_s^2}f_{u,\alpha} + \frac{\mathcal{H}^{(2)}_{i,\alpha\beta}}{2c_s^4}a_{\alpha\beta}^{F,(2)} \right],
\label{eq_fi}
\end{equation}
where $a_{\alpha\beta}^{F,(2)}$ is the Hermite moment defined in Eq. (\ref{eq_af2}) of \ref{app_d3q19}.

After solving Eq. (\ref{eq_LB}), the macroscopic quantities $\rho$ and $\boldsymbol{u}$ are obtained as

\begin{equation}
    \rho = \sum_i \bar{f}_i,
\end{equation}

\begin{equation}
    \rho \boldsymbol{u}= \sum_i \boldsymbol{c}_{i}\bar{f}_i + \frac{\Delta t}{2}\boldsymbol{f}_{\boldsymbol{u}},
\end{equation}
where the momentum forcing term $\boldsymbol{f}_{\boldsymbol{u}}=\boldsymbol{f}^\mathrm{IB}$ is defined in Eq. (\ref{eq_fib2}) of Section \ref{sub_recons}.

To facilitate the application of the hybrid LBM for compressible flows, Wissocq et al. \cite{wissocq2022restoring} proposed a scheme that couples the LBM with the total energy equation. The resulting discrete form of the energy equation, incorporating the forcing term, is expressed as \cite{wissocq2022restoring}:

\begin{equation}
\delta_t(\rho E)+\delta_\alpha F_{+\Delta\alpha/2}^{\rho E, NS}=f_{E},
\label{eq_energy}
\end{equation}
where $\delta_t$ and $\delta_\alpha$ denote discrete operators defined in Eq (\ref{eq_deltata}) of \ref{app_d3q19}. The flux $F_{+\Delta\alpha/2}^{\rho E, NS}$ is expressed as:
\begin{equation}
\begin{aligned}
F_{+\Delta\alpha/2}^{\rho E, NS} &= \mathscr{F}_{+\Delta\alpha/2}^*(\rho Hu_\alpha)+(h-k)\left[F_{+\Delta\alpha/2}^\rho-\mathscr{F}_{+\Delta\alpha/2}^*(\rho u_\alpha)\right] \\
&+u_\beta\left[F_{+\Delta\alpha/2}^{\rho u_\beta}-\mathscr{F}_{+\Delta\alpha/2}^*(\rho u_\alpha u_\beta+p\delta_{\alpha\beta})\right] \\
&-\lambda\delta_\alpha T(\boldsymbol{x}+\boldsymbol{e}_\alpha\Delta x,t),
\end{aligned}
\end{equation}
where $H=E+p/\rho$ is the total enthalpy. $h=\gamma RT/(\gamma-1)$ is the enthalpy and $k=u_\alpha^2/2$ is the kinetic energy. $\boldsymbol{e}_\alpha$ denotes the unity vector in the direction $\alpha$. $F_{+\Delta\alpha/2}^\rho$ and $F_{+\Delta\alpha/2}^{\rho u}$ are mass and momentum fluxes defined in Eqs. (\ref{eq_fluxx}), (\ref{eq_fluxy}) and (\ref{eq_fluxz}) of \ref{app_d3q19}. $\mathscr{F}_{+\Delta\alpha/2}^*$ is the linear function defined in Eq. (\ref{eq_linear}) of \ref{app_d3q19}, determined using the MUSCL-Hancock scheme \cite{wissocq2022restoring}. The energy forcing term $f_{E}$ including the kinetic energy part $w^\mathrm{IB}$ and the internal energy part $q^\mathrm{IB}$ is defined in Eqs. (\ref{eq_wib2}) and (\ref{eq_qib2}) of Section \ref{sub_recons}. After solving the Eq. (\ref{eq_energy}), the temperature is obtained from $E=||\boldsymbol{u}||^2/2+C_v T$ and the pressure can be calculated by $p=\rho R T$.

\subsection{Fully one-sided diffuse-interface IBM}
\label{sub_IB}

\begin{figure}
\centerline{\includegraphics[width=0.7\linewidth]{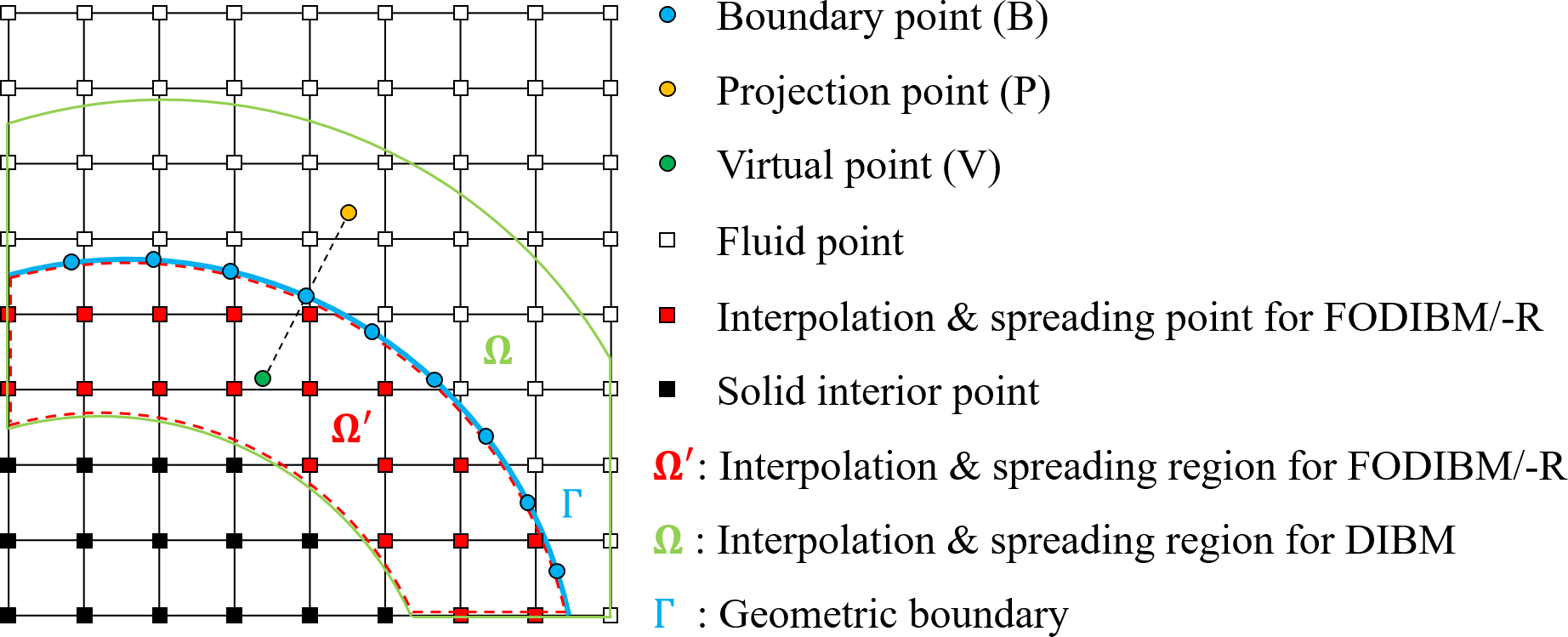}}
\caption{Schematic of the FODIBM-R. Each boundary point is associated with a projection point located at a distance $d_n$ along the interface normal. This projection point is used to impose the Neumann boundary condition and reconstruct the target values of the flow variables at the virtual point. The position of the virtual point is determined by the geometry.}
\label{fig_one-side}
\end{figure}

In the present study, several immersed boundary methods are considered for comparison. For brevity, their abbreviations are summarized in Table \ref{table_abb}. Figure \ref{fig_one-side} illustrates the proposed immersed boundary method. The geometric boundary ($\Gamma$) is discretized by a set of Lagrangian points ($\boldsymbol{X}_l$) defined in a curvilinear coordinate system $(q,r,s)$. The fluid domain is discretized on a fixed Cartesian mesh. The set $\Omega$ denotes all Eulerian points ($\boldsymbol{x}_i$) located both inside and outside the geometric boundary, while $\Omega^{\prime}$ represents the subset of Eulerian points located inside the boundary. In the traditional diffuse-interface IBM (DIBM), both interpolation and spreading operations are performed over the entire set $\Omega$, i.e. on grid points located on both sides of the boundary. This treatment introduces a diffusion effect across the interface, which degrades the accuracy of the near-boundary flow field and hinders the proper enforcement of Neumann boundary conditions. In contrast, the fully one-sided diffuse-interface IBM (FODIBM) restricts both interpolation and spreading to $\Omega^{\prime}$. This one-sided treatment effectively eliminates the diffusion-related issues inherent in the DIBM. As a result, the FODIBM exhibits significantly improved performance, particularly for compressible flows \cite{mao2026explicit}. The projection point is used to impose the Neumann boundary condition and reconstruct the target values of the flow variables at the virtual point. Both concepts are introduced later.

\begin{table}[t]
\centering
\small
\begin{tabular}{l c}
  \hline
  Method & Abbreviation \\
  \hline
  Diffuse-interface immersed boundary method & DIBM \\
  Fully one-sided diffuse-interface immersed boundary method & FODIBM \\
  FODIBM with target-value reconstruction & FODIBM-R \\
  Sharp-interface immersed boundary method & SIBM \\
  \hline
\end{tabular}
\caption{Abbreviations of different immersed boundary methods.}
\label{table_abb}
\end{table}

We first present the detailed formulations of both the DIBM and the FODIBM. The IB forcing is computed from the predicted flow variables interpolated at the Lagrangian points. The resulting forcing term is then incorporated into the governing equations by spreading the IB force onto the surrounding Eulerian points. We begin with the DIBM, in which the interpolation step is formulated as

\begin{equation}
    \boldsymbol{M}^*_l=\mathcal{I}[\boldsymbol{m}]_l=\sum_{\Omega} \boldsymbol{m}_i\delta(\boldsymbol{x}_{i}-\boldsymbol{X}_{l})\Delta x^3,
\label{eq_int1}
\end{equation}
where $\boldsymbol{x}_i \in \Omega$ indicates the fluid points both inside and outside the geometric boundary. $\boldsymbol{m}$ denotes flow variables at Eulerian points, such as velocity, density and temperature. $\boldsymbol{M}^*$ is the predicted flow variables at Lagrangian points. $\mathcal{I}[\bullet]_l$ is the interpolation operator. The spreading step is expressed as

\begin{equation}
    \boldsymbol{g}_i^{\mathrm{IB}}=\mathcal{S}[\boldsymbol{G}^{\mathrm{IB}}]_i=\sum_{\Gamma} \boldsymbol{G}_l^{\mathrm{IB}}\delta(\boldsymbol{x}_{i}-\boldsymbol{X}_{l})W_l,
\label{eq_spr1}
\end{equation}
where $\boldsymbol{X}_l \in \Gamma$. $\boldsymbol{g}^{\mathrm{IB}}$ is the IB forcing term which will be incorporated into the governing equations, such as the momentum forcing term ($\boldsymbol{f}^{\mathrm{IB}}$) and the energy forcing term (includes kinetic energy part $w^{\mathrm{IB}}$ and internal energy part $q^{\mathrm{IB}}$). $\boldsymbol{G}^{\mathrm{IB}}$ (includes $\boldsymbol{F}^{\mathrm{IB}}$, $W^{\mathrm{IB}}$ and ${Q}^{\mathrm{IB}}$) is the IB force at Lagrangian points. $W_l=\Delta q \Delta r \Delta s$ is the Lagrangian weight. An improved Lagrangian weight recovering the interpolation/spreading reciprocity condition will be defined hereafter, which decreases errors in both Dirichlet and Neumann boundary conditions \cite{mao2026explicit}.

For the FODIBM, both the interpolation and spreading steps are corrected by introducing a scaling factor ($\phi_l$) to restore the equivalence between the Eulerian and Lagrangian variables, i.e. the conservation of physical quantities such as flow variables, forces, and momentum \cite{mao2026explicit}. This correction is necessary because the interpolation domain is reduced from $\Omega$ to $\Omega^{\prime}$:

\begin{equation}
    \boldsymbol{M}^*_l=\mathcal{I}[\boldsymbol{m}]_l=\sum_{\Omega ^ {\prime}} \phi_l\boldsymbol{m}_i\delta(\boldsymbol{x}_{i}-\boldsymbol{X}_{l})\Delta x^3,
\label{eq_int2}
\end{equation}

\begin{equation}
    \boldsymbol{g}_i^{\mathrm{IB}}=\mathcal{S}[\boldsymbol{G}^{\mathrm{IB}}]_i=\sum_{\Gamma} \phi_l\boldsymbol{G}_l^{\mathrm{IB}}\delta(\boldsymbol{x}_{i}-\boldsymbol{X}_{l})W_l,
\label{eq_spr2}
\end{equation}
where $\phi_l$ is defined as

\begin{equation}
\phi_l=\frac{1}{\sum_{\Omega ^ {\prime}}\delta(\boldsymbol{x}_{i}-\boldsymbol{X}_{l})\Delta x^3}.
\label{eq_phi}
\end{equation}

For the conventional direct-forcing IBM, the interpolation/spreading is non-reciprocal ($\mathcal{I}[\mathcal{S}[\boldsymbol{G}^{\mathrm{IB}}]]\neq\boldsymbol{G}^{\mathrm{IB}}$) in explicit implementations, which causes a boundary error as demonstrated by Gsell and Favier \cite{gsell2019explicit}. This error corresponds to discrepancies between the target and the realized values at the boundary. To enforce reciprocity between interpolation and spreading, an improved Lagrangian weight is introduced, such that $\mathcal{I}[\mathcal{S}[\boldsymbol{G}^{\mathrm{IB}}]]\approx \boldsymbol{G}^{\mathrm{IB}}$. In other words, spreading $\boldsymbol{G}^{\mathrm{IB}}$ onto the Eulerian field and then interpolating it back yields approximately the same value. The improved Lagrangian weight ($W_l$) normalized by $\Delta x^3$ is defined as

\begin{equation}
    W_l = \frac{1}{\sum_i\sum_k[\phi_k\delta(\boldsymbol{x}_i-\boldsymbol{X}_k)][\phi_l\delta(\boldsymbol{x}_i-\boldsymbol{X}_l)]}.
\label{eq_weight}
\end{equation}

The derivation of $W_l$ is presented in our previous work \cite{mao2026explicit}. The Lagrangian weight is determined exclusively by the geometry of the boundary and its discretization. The associated computational cost is low because it only depends on a small number of neighboring Eulerian and Lagrangian points.

In summary, the value of $\phi_l$ in Eqs. (\ref{eq_int2}), (\ref{eq_spr2}), and (\ref{eq_weight}) is set to unity for the DIBM, whereas it is computed using Eq. (\ref{eq_phi}) for the FODIBM.

The delta function ($\delta$) for interpolation and spreading is expressed as

\begin{equation}
\delta(\boldsymbol{x}_{i}-\boldsymbol{X}_{l})=\frac{1}{\Delta x^3}\widetilde{\delta}\left(\frac{{x}_{i}-X_{l}}{\Delta x}\right)\widetilde{\delta}\left(\frac{{y}_{i}-Y_{l}}{\Delta x}\right)\widetilde{\delta}\left(\frac{{z}_{i}-Z_{l}}{\Delta x}\right),
\end{equation}

\begin{equation}
\widetilde{\delta}(r)=\begin{cases}\frac{1}{2d}[1+\cos(\frac{\pi r}{d})] &|r|\leq d,\\0 &|r|>d,\end{cases}
\end{equation}
where $d=2$ is the radius of the delta function.

To enforce the desired boundary conditions, the IB forcing terms $\boldsymbol{g}^{\mathrm{IB}}$, including the momentum forcing term $\boldsymbol{f}^{\mathrm{IB}}$ and the energy forcing terms ${w}^{\mathrm{IB}}$ and ${q}^{\mathrm{IB}}$, are introduced. The momentum forcing term at a Lagrangian point is expressed as:

\begin{equation}
    \boldsymbol{F}^{\mathrm{IB}}_l=\frac{2[\rho^*\boldsymbol{u}^t- (\rho\boldsymbol{u})^*]}{\Delta t},
\label{eq_fib}
\end{equation}
where $\boldsymbol{u}^t$ is the target velocity at a Lagrangian point, determined by the prescribed boundary condition. For example, $\boldsymbol{u}^t$ is set to zero for a no-slip boundary condition. The quantities $\rho^*$ and $(\rho \boldsymbol{u})^*$ denote the predicted density and momentum obtained from Eq. (\ref{eq_int2}). The Eulerian forcing term $\boldsymbol{f}^{\mathrm{IB}}$ is then obtained by spreading $\boldsymbol{F}^{\mathrm{IB}}$ to the surrounding Eulerian points using Eq. (\ref{eq_spr2}).

The energy forcing term at a Lagrangian point can be decomposed into two parts:
\begin{equation}
    {W}^{\mathrm{IB}}_l=\frac{\rho^*[(\boldsymbol{u}^t)^2- (\boldsymbol{u}^*)^2]}{2\Delta t},
\label{eq_wib}
\end{equation}

\begin{equation}
    {Q}^{\mathrm{IB}}_l=\frac{C_v\rho^*(T^t- T^*)}{\Delta t},
\label{eq_qib}
\end{equation}
where $\boldsymbol{u}^*$ and $T^*$ denote the predicted velocity and temperature obtained from Eq. (\ref{eq_int2}). The target temperature $T^t$ at a Lagrangian point is determined by the prescribed boundary condition. For an isothermal boundary condition, $T^t$ is set to a constant for all Lagrangian points. For an adiabatic boundary condition ($\partial T/\partial n = 0$), a projection point is defined at a distance $d_n$ along the interface normal from each Lagrangian point, as illustrated in Fig. \ref{fig_one-side}. To satisfy $\partial T/\partial n = 0$, the target temperature $T^t$ is set equal to the temperature at the projection point, denoted by $T^{P*}$. The value of $T^{P*}$ is obtained by interpolation using Eq. (\ref{eq_int1}), where the radius of the delta function is set to $d = 1$ to prevent the inclusion of information from inside the solid. In the present study, $d_n/\Delta x$ is set to 1.5 to ensure accurate enforcement of the Neumann boundary condition, as demonstrated in our previous study \cite{mao2026explicit}. After obtaining ${W}^{\mathrm{IB}}$ and ${Q}^{\mathrm{IB}}$, the corresponding Eulerian forcing terms ${w}^{\mathrm{IB}}$ and ${q}^{\mathrm{IB}}$ are computed by spreading ${W}^{\mathrm{IB}}$ and ${Q}^{\mathrm{IB}}$ to the surrounding Eulerian points using Eq. (\ref{eq_spr2}).

\subsection{Target-value reconstruction}
\label{sub_recons}

The initial motivation for investigating the influence of the one-sided spreading scheme on boundary representation arose from a phenomenon observed in previous work \cite{mao2026explicit}. In particular, the shock stand-off distance of the bow shock predicted by the original FODIBM was smaller than the reference results. First, this issue is not related to the conservative property of the interpolation/spreading process, since the scaling factor ($\phi_l$) compensates for the loss of symmetry in the interpolation/spreading region and restores the equivalence between the Eulerian and Lagrangian variables, including flow variables, forces, and momentum \cite{mao2026explicit}:

\begin{equation}
\sum_{\Omega ^ {\prime}} \boldsymbol{m}_i\Delta x^3 = \sum_{\Gamma} \boldsymbol{M}_l^*\Delta q\Delta r\Delta s,
\label{eq_equiv}
\end{equation}
where $\boldsymbol{M}_l^*$ is calculated by Eq. (\ref{eq_int2}).

For the DIBM, the geometric boundary position $\boldsymbol{X}_{l}$ and the effective boundary position $\boldsymbol{X}_{l}^*$ obtained by interpolation are equivalent due to the symmetry of the interpolation/spreading region:

\begin{equation}
    \boldsymbol{X}_{l}=\boldsymbol{X}^*_l=\sum_{\Omega} \boldsymbol{x}_i\delta(\boldsymbol{x}_{i}-\boldsymbol{X}_{l})\Delta x^3,
\end{equation}
which indicates that the boundary condition is enforced at $\boldsymbol{X}_{l}$ ($\boldsymbol{X}_{l}^*$) when spreading the forcing term to both sides of the boundary. The delta function ensures a smooth transition of the flow variables near the boundary, allowing them to reach the desired values at $\boldsymbol{X}_{l}$ ($\boldsymbol{X}_{l}^*$).

In contrast, for the FODIBM, the mismatch between $\boldsymbol{X}_{l}$ and $\boldsymbol{X}_{l}^*$ arises from the loss of symmetry in the spreading region. When the forcing term is spread only inside the boundary, $\boldsymbol{X}_{l}^*$ becomes:

\begin{equation}
    \boldsymbol{X}^*_l=\sum_{\Omega^{\prime}} \phi_l \boldsymbol{x}_i\delta(\boldsymbol{x}_{i}-\boldsymbol{X}_{l})\Delta x^3.
\label{eq_xeffect}
\end{equation}

Although $\phi_l$ restores the conservative property of the interpolation/spreading process in Eq. (\ref{eq_equiv}), the effective boundary position $\boldsymbol{X}_{l}^*$ shifts inward because it represents the average of $\boldsymbol{x}_i$ over nearby interior Eulerian points. In other words, the boundary condition is applied to a position inside the geometric boundary.

\begin{figure}
\centerline{\includegraphics[width=1.0\linewidth]{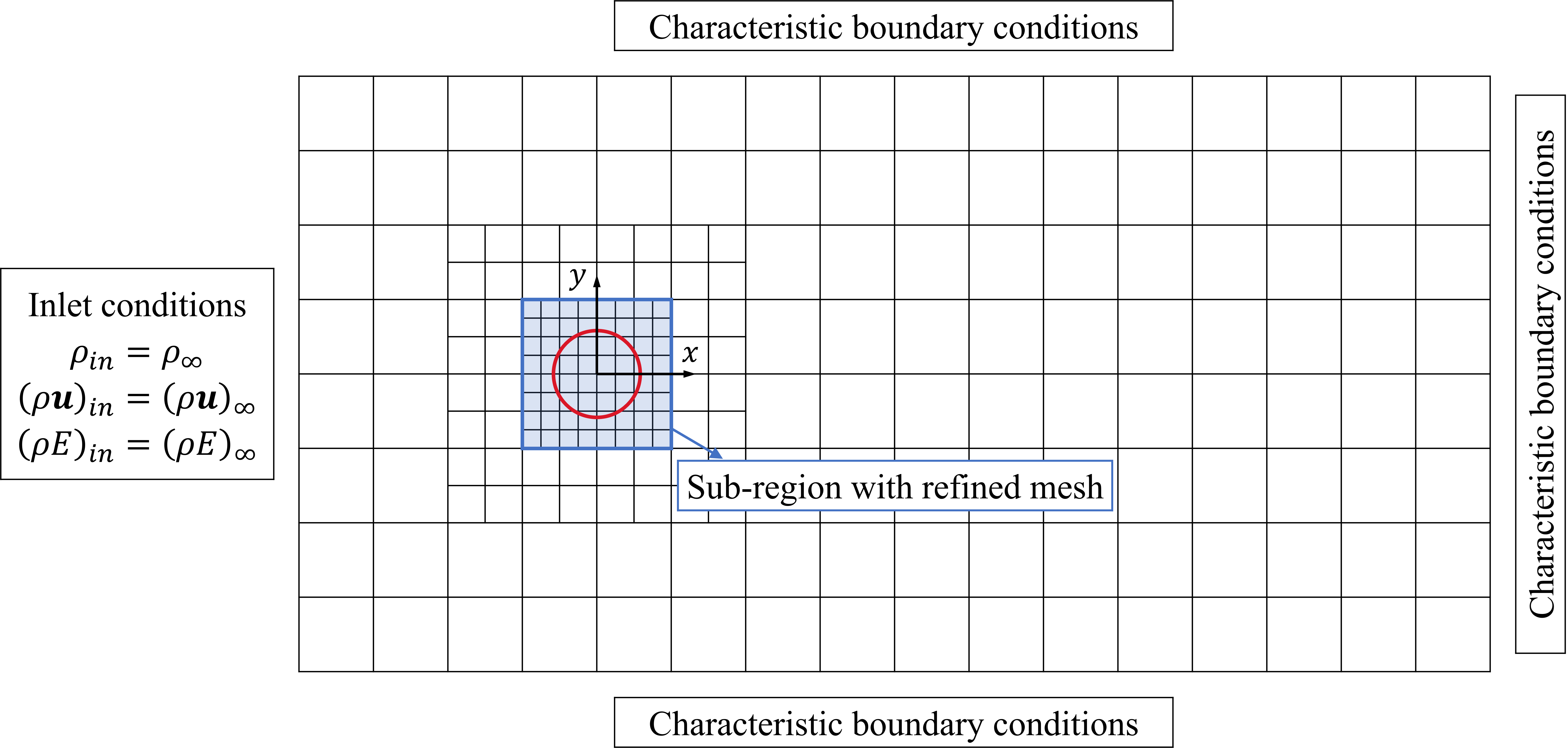}}
\caption{Schematic of the computational mesh and boundary conditions.}
\label{fig_schedomain}
\end{figure}

\begin{figure}
\centerline{\includegraphics[width=0.5\linewidth]{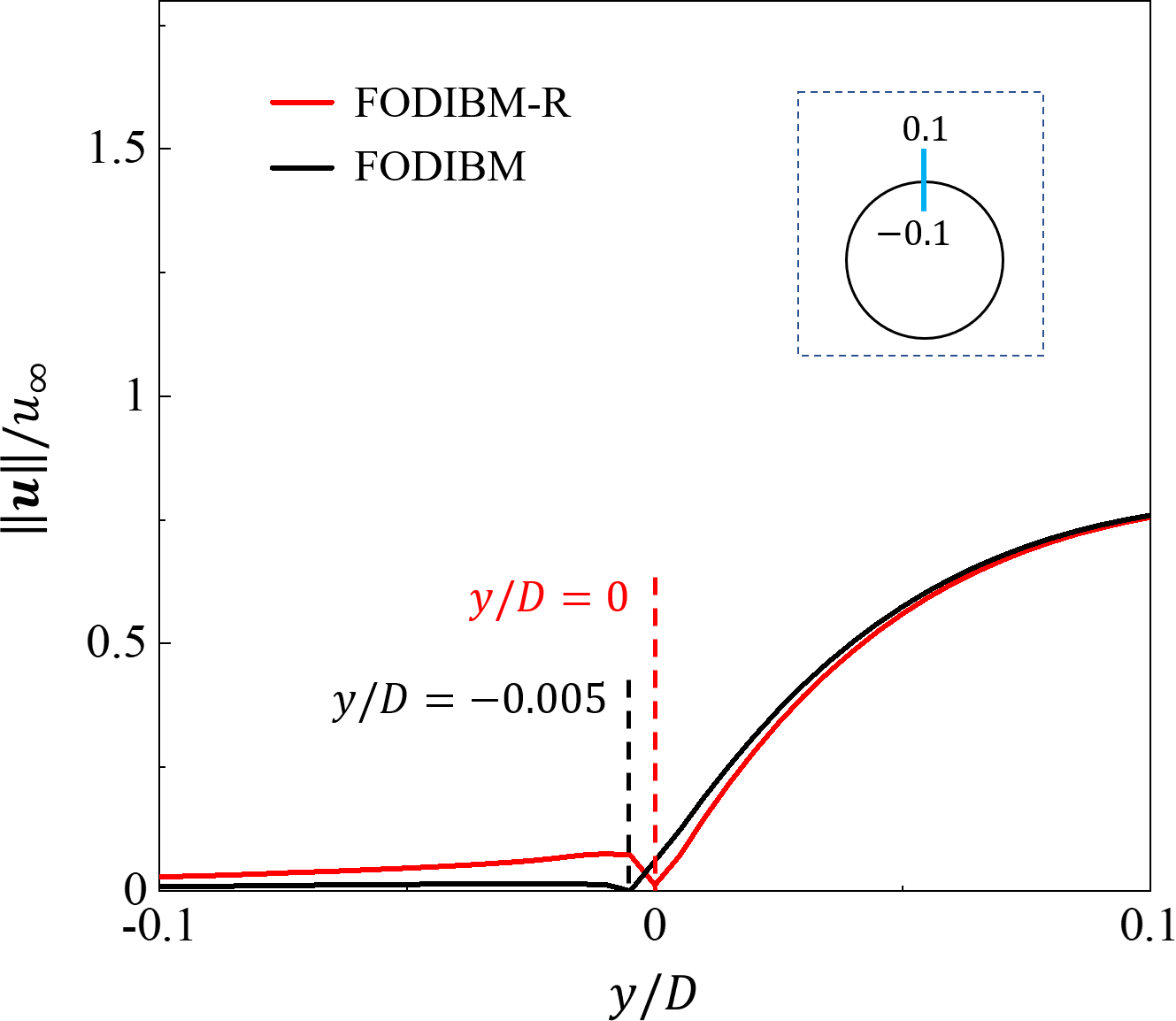}}
\caption{Velocity profiles near the wall for supersonic flow past an adiabatic circular cylinder. The data extraction line is indicated in blue (not to scale).}
\label{fig_wallposition}
\end{figure}

To demonstrate this, we simulate supersonic flow past an adiabatic circular cylinder of diameter $D$. Figure \ref{fig_schedomain} shows a schematic of the computational mesh and boundary conditions. All conservative variables are prescribed at the inlet, while characteristic boundary conditions \cite{poinsot1992boundary} are applied at the top, bottom, and outlet boundaries to enforce non-reflecting behavior. The cylinder center is located at (0,0) inside a rectangular domain of dimensions [-10$D$, 30$D$]$\times$[-10$D$, 10$D$]. The mesh spacing in the refined sub-region [-2$D$, 2$D$]$\times$[-2$D$, 2$D$] is $\Delta x_{\mathrm{min}}=D/200$, with a fixed Eulerian-to-Lagrangian mesh ratio of $\Delta x/\Delta s = 1$. The free-stream Mach and Reynolds numbers are set to $Ma_{\infty}=u_{\infty}/c_{\infty}=2.0$ and $Re=\rho_{\infty}u_{\infty}D/\mu=300$, respectively, where $u_{\infty}$, $c_{\infty}$, $\rho_{\infty}$, and $\mu$ denote the free-stream velocity, sound speed, density, and dynamic viscosity. No-slip ($\boldsymbol{u}^t=0$) and adiabatic ($\partial T/\partial n=0$) conditions are applied on the geometric boundary. Fig. \ref{fig_wallposition} shows the velocity distribution near the cylinder surface. The data extraction line is indicated in blue, with $y/D=0$ representing the defined boundary position $\boldsymbol{X}_{l}$. For the FODIBM result, the location of minimal velocity shifts slightly from $y/D=0$ to $y/D=-0.005$, which coincides with $\boldsymbol{X}_{l}^*$ obtained from Eq. (\ref{eq_xeffect}), confirming the above analysis.

The inward shift of $\boldsymbol{X}_{l}^*$ is inherent to the one-sided spreading scheme and cannot be directly corrected to coincide with $\boldsymbol{X}_{l}$. In particular, shifting the Lagrangian point outward is not feasible, as it would lead to the spreading of IB forcing into the fluid region. To address this issue, a target-value reconstruction strategy is proposed. The virtual point $V$, geometric boundary point $B$, and projection point $P$ involved in the reconstruction are illustrated in Fig. \ref{fig_one-side}. For each Lagrangian point (point $B$), a virtual point $V$ is defined at $\boldsymbol{X}_{l}^*$. Its location is determined quantitatively from Eq. (\ref{eq_xeffect}), rather than prescribed through an empirical parameter. The key idea is to reconstruct the target values of the flow variables at the virtual point $V$ ($\boldsymbol{M}^{Vt}$) using the predicted values at the projection point $P$ ($\boldsymbol{M}^{P*}$) and the prescribed target values at the boundary point $B$ ($\boldsymbol{M}^{Bt}$). The reconstructed target values are then used to update the corresponding forcing terms, which is subsequently spread to the Eulerian grid. In this manner, the desired boundary conditions are indirectly enforced at the boundary point $B$.

For Dirichlet boundary conditions, $\boldsymbol{M}^{Vt}$ is obtained via linear interpolation along the normal direction:

\begin{equation}
    \frac{\boldsymbol{M}^{P*}-\boldsymbol{M}^{Bt}}{d_{PB}}=\frac{\boldsymbol{M}^{Bt}-\boldsymbol{M}^{Vt}}{d_{BV}},
\end{equation}

\begin{equation}
    \boldsymbol{M}^{Vt}=\boldsymbol{M}^{Bt}-\frac{d_{BV}}{d_{PB}}(\boldsymbol{M}^{P*}-\boldsymbol{M}^{Bt}),
\label{eq_virtualdir}
\end{equation}
where $d_{BV}$ and $d_{PB}$ denote the distances between points $B$-$V$ and $P$-$B$, respectively.

For Neumann boundary conditions, $\boldsymbol{M}^{Vt}$ is given by

\begin{equation}
    \frac{\partial \boldsymbol{M}}{\partial n}=\frac{\boldsymbol{M}^{P*}-\boldsymbol{M}^{Vt}}{d_{PB}+d_{BV}},
\end{equation}

\begin{equation}
    \boldsymbol{M}^{Vt}=\boldsymbol{M}^{P*}- \frac{\partial \boldsymbol{M}}{\partial n}(d_{PB}+d_{BV}).
\label{eq_virtualneu}
\end{equation}

In particular, for Neumann boundary conditions with zero gradient (e.g. adiabatic walls), Eq. (\ref{eq_virtualneu}) reduces to $\boldsymbol{M}^{Vt}=\boldsymbol{M}^{P*}$.

Although the reconstruction is formally similar to that used in ghost-point methods, the present approach avoids explicit identification of Eulerian ghost points and the associated geometric construction of boundary-intersection and image points. Instead, the virtual-point location is determined directly from the effective boundary position given by Eq. (\ref{eq_xeffect}).

In summary, for the FODIBM with target-value reconstruction (FODIBM-R), the target velocity $\boldsymbol{u}^{Vt}$ and temperature $T^{Vt}$ are evaluated using Eq. (\ref{eq_virtualdir}) or Eq. (\ref{eq_virtualneu}), depending on the prescribed boundary conditions. The corresponding forcing terms defined in Eqs. (\ref{eq_fib}), (\ref{eq_wib}), and (\ref{eq_qib}) are then modified as

\begin{equation}
    \boldsymbol{F}^{\mathrm{IB}}_l=\frac{2[\rho^*\boldsymbol{u}^{Vt}- (\rho\boldsymbol{u})^*]}{\Delta t},
\label{eq_fib2}
\end{equation}

\begin{equation}
    {W}^{\mathrm{IB}}_l=\frac{\rho^*[(\boldsymbol{u}^{Vt})^2- (\boldsymbol{u}^*)^2]}{2\Delta t},
\label{eq_wib2}
\end{equation}

\begin{equation}
    {Q}^{\mathrm{IB}}_l=\frac{C_v\rho^*(T^{Vt}- T^*)}{\Delta t},
\label{eq_qib2}
\end{equation}

As shown in Fig. \ref{fig_wallposition}, the proposed FODIBM-R correctly restores the location of minimum velocity to $y/D=0$, thereby providing preliminary validation of its effectiveness.

\subsection{Computational procedure}
\label{sub_Algorithm}

The computational procedure for each time step is summarized as follows:

1. Compute the predicted variables $\rho^*$, $\boldsymbol{u}^*$, and $T^*$ using Eq. (\ref{eq_int2}). The flow variables from the previous time step are treated as intermediate variables;

2. Evaluate $\boldsymbol{u}^{P*}$ and $T^{P*}$ at the projection points using Eq. (\ref{eq_int1});

3. Compute the IB forcing terms at all Lagrangian points: for the FODIBM-R using Eqs. (\ref{eq_fib2})-(\ref{eq_qib2}), and for the FODIBM using Eqs. (\ref{eq_fib})-(\ref{eq_qib});

4. Spread the IB forcing terms to the surrounding Eulerian points using Eq. (\ref{eq_spr2});

5. Solve Eqs. (\ref{eq_LB}) and (\ref{eq_energy}) with the IB forcing terms to update $\rho$, $\boldsymbol{u}$, $T$, and $p$.

\section{Results and discussions}
\label{res}

In this section, we first compare the numerical convergence of the $\mathrm{L}_2$ and $\mathrm{L}_{\infty}$ errors between the FODIBM-R and the FODIBM. The errors associated with Dirichlet and Neumann boundary conditions are then evaluated for both methods. The computational cost of the FODIBM-R is subsequently examined and compared with that of the FODIBM. Finally, the effectiveness of the FODIBM-R for compressible flows around complex geometries is demonstrated through a series of validation cases.

\subsection{Numerical convergence of error norms}
\label{sub_conv}

\begin{figure}
\centerline{\includegraphics[width=1.0\linewidth]{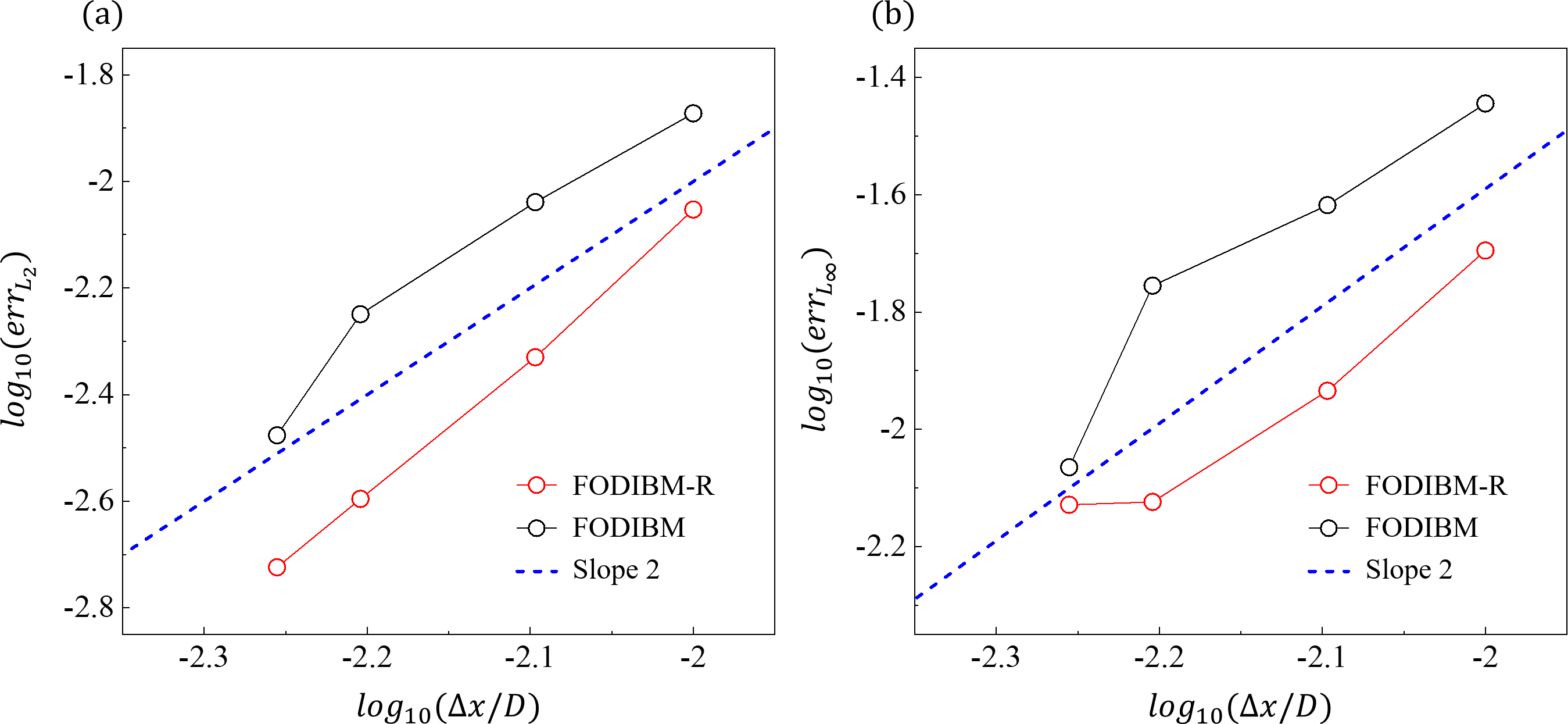}}
\caption{Numerical convergence of the (a) $\mathrm{L}_2$ and (b) $\mathrm{L}_{\infty}$ errors for supersonic flow past an adiabatic circular cylinder.}
\label{fig_conv}
\end{figure}

To assess the overall accuracy of the present frameworks employing the FODIBM-R and FODIBM, a series of simulations of supersonic flow past an adiabatic circular cylinder is conducted using five mesh resolutions, namely $\Delta x_{\mathrm{min}} = D/100$, $D/125$, $D/160$, $D/180$, and $D/200$. The computational setup is identical to that described in Section \ref{sub_recons}. The $\mathrm{L}_2$ and $\mathrm{L}_{\infty}$ errors of the pressure coefficient ($C_p = [p - p_{\infty}] / [0.5 \rho_{\infty} u_{\infty}^2]$) are defined as

\begin{equation}
err_{L_2} = \sqrt{
\frac{\sum_{i=1}^{N_L} \left[ (C_p)_i - (C_p)_i^{\mathrm{ref}} \right]^2}
{\sum_{i=1}^{N_L} \left[ (C_p)_i^{\mathrm{ref}} \right]^2}
},
\end{equation}

\begin{equation}
err_{L_\infty} = \max_{1 \le i \le N_L} \left| (C_p)_i - (C_p)_i^{\mathrm{ref}} \right|.
\end{equation}

Here, $(C_p)^{\mathrm{ref}}$ is obtained from the reference solution computed on the finest mesh ($\Delta x_{\mathrm{min}} = D/200$). $N_L$ represents the number of Lagrangian points. For all cases, the pressure values are interpolated from the Eulerian field at identical spatial locations.

The convergence behavior is illustrated in Fig. \ref{fig_conv}. In terms of the $\mathrm{L}_2$ error, both frameworks exhibit approximately second-order spatial accuracy, indicating that neither the FODIBM-R nor the FODIBM degrades the overall accuracy of the LBM. This behavior can be attributed to the use of the scaling factor and the improved Lagrangian weight, which restore the conservative property and reciprocity of the interpolation and spreading operations, thereby reducing discretization errors. For the $\mathrm{L}_{\infty}$ error, the convergence rate is less pronounced. Nevertheless, the FODIBM-R consistently yields lower error levels across all mesh resolutions compared to the FODIBM. This improvement is mainly attributed to a more accurate representation of the boundary location, which leads to enhanced prediction.

\subsection{Boundary condition error analysis}
\label{sub_bcerr}

\begin{figure}
\centerline{\includegraphics[width=1.0\linewidth]{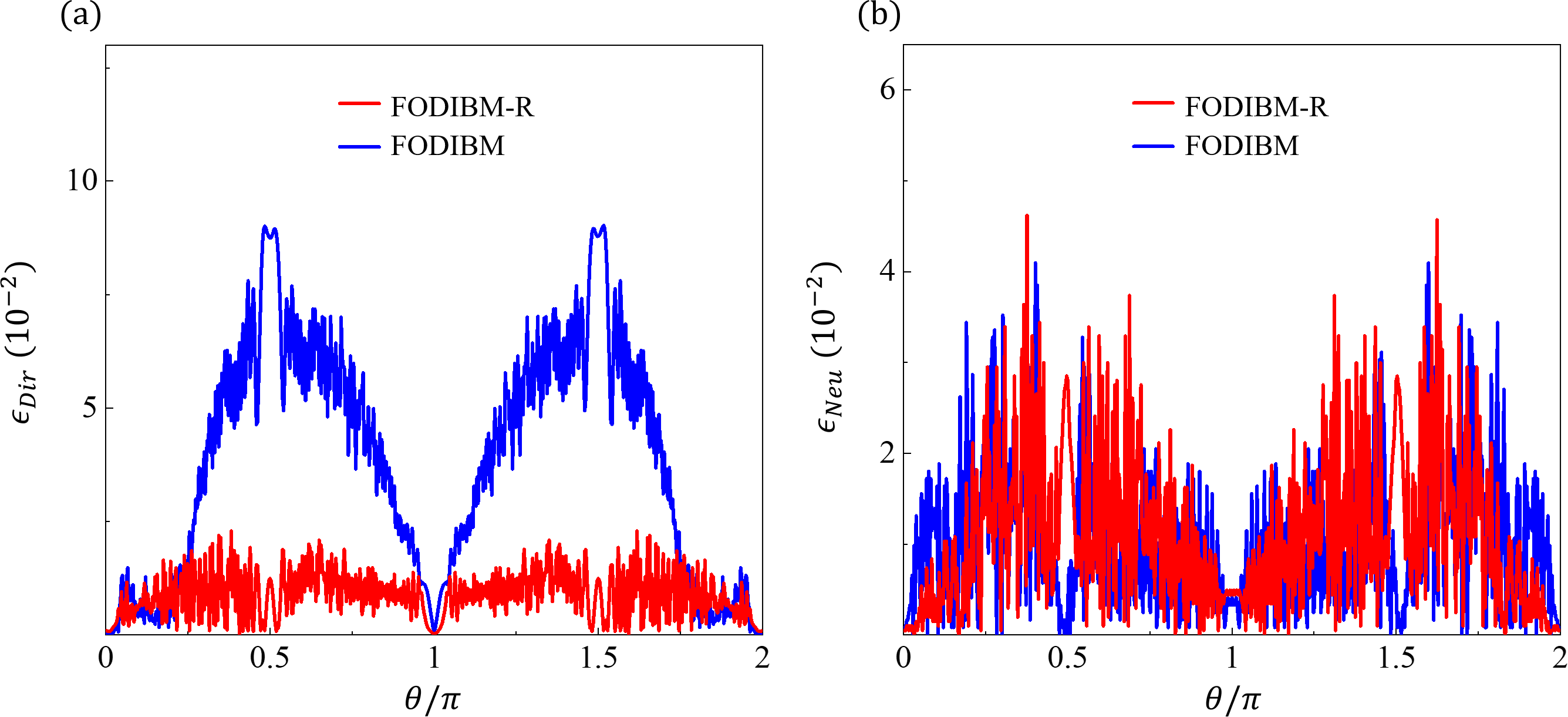}}
\caption{Distribution of boundary errors on an adiabatic circular cylinder for FODIBM-R and FODIBM: (a) Dirichlet condition; (b) Neumann condition ($\theta/\pi = 0$ and $1$ denote the leading and trailing edges, respectively).}
\label{fig_error}
\end{figure}

\begin{table}[t]
\centering
\small
\begin{tabular}{l c c c c}
  \hline
    & $\bar{\epsilon}_{Dir}$ & $\Delta\epsilon~(\%)$ & $\bar{\epsilon}_{Neu}$ & $\Delta\epsilon~(\%)$ \\
  \hline
  FODIBM-R & $8.90 \times 10^{-3}$ & $-77$ & $1.10 \times 10^{-2}$ & $+8$ \\
  FODIBM & $3.81 \times 10^{-2}$ & - & $1.02 \times 10^{-2}$ & - \\
  \hline
\end{tabular}
\caption{Averaged Dirichlet ($\bar{\epsilon}_{Dir}$) and Neumann ($\bar{\epsilon}_{Neu}$) boundary errors on an adiabatic circular cylinder for FODIBM-R and FODIBM.}
\label{table_error}
\end{table}

Here, we further compare the Dirichlet and Neumann boundary condition errors for the FODIBM-R and the FODIBM in the case of supersonic flow past an adiabatic circular cylinder. The computational configuration follows that described in Section \ref{sub_recons}. The no-slip (Dirichlet) boundary condition error is defined as

\begin{equation}
{\epsilon}_{Dir}=\frac{|u^s-u^t|}{u_{\infty}},
\end{equation}
where the superscript $s$ denotes variables interpolated from the updated flow field using Eq. (\ref{eq_int1}), and $t$ denotes the prescribed target values on the geometric boundary.

The adiabatic (Neumann) boundary condition error is defined as
\begin{equation}
{\epsilon}_{Neu}=\frac{|(\frac{\partial T}{\partial n})^s-(\frac{\partial T}{\partial n})^t|}{T_{\infty}/D}.
\end{equation}

Figure \ref{fig_error} shows the distribution of boundary errors on the cylinder surface. The no-slip boundary condition error of the FODIBM-R is significantly reduced compared to that of the FODIBM. This improvement is expected, as the effective boundary position shifts inward in the FODIBM but is corrected in the FODIBM-R. For the adiabatic boundary condition, both methods exhibit similar error levels. This is because the reconstructed target temperature in the FODIBM-R is consistent with that used in the FODIBM, as discussed in Section \ref{sub_recons}. To provide a quantitative comparison, the space-averaged Dirichlet and Neumann boundary condition errors are reported in Table \ref{table_error}. Compared with the FODIBM, the FODIBM-R reduces $\bar{\epsilon}_{Dir}$ to $8.9 \times 10^{-3}$, corresponding to a decrease of $77\%$, while $\bar{\epsilon}_{Neu} = 1.1 \times 10^{-2}$ shows a slight increase of $8\%$. These results demonstrate that the FODIBM-R significantly improves the enforcement of the Dirichlet boundary condition, while maintaining accurate enforcement of the Neumann boundary condition.

\subsection{Computational cost analysis}

The above results indicate that the FODIBM-R achieves a consistent reduction in error norms across different grid resolutions, along with a significant improvement in boundary condition accuracy. However, evaluating the computational cost is also essential, as an additional reconstruction step is introduced in the FODIBM-R. Here, the computational time cost of the FODIBM-R and the FODIBM is examined over 1000 time steps for the case of supersonic flow past an adiabatic circular cylinder. The computational configuration follows that described in Section \ref{sub_recons}. Table \ref{table_cost} summarizes the computational time cost for different numbers of Lagrangian points ($N_L$). As $N_L$ increases from 500 ($\Delta x = \Delta s = D/160$) to 5000 ($\Delta x = \Delta s = D/1600$), the computational time of the FODIBM-R is consistently higher than that of the FODIBM. However, the increase remains marginal, with a maximum increment of only $1.408\%$. These results demonstrate that the target-value reconstruction based on linear interpolation introduces only negligible additional computational cost, while the FODIBM-R provides a substantial improvement in accuracy, achieving a favorable balance between accuracy and efficiency.

\begin{table}[t]
\centering
\small
\begin{tabular}{l c c c c}
  \hline
  Number of Lagrangian points & 500 & 1000 & 2500 & 5000 \\
  \hline
  FODIBM-R & 20.622 s & 48.620 s & 251.307 s & 1056.172 s \\
  FODIBM & 20.540 s & 47.947 s & 248.212 s & 1041.503 s \\
  $\Delta (\%)$ & 0.399 & 1.403 & 1.246 & 1.408 \\
  \hline
\end{tabular}
\caption{Computational time cost of the FODIBM-R and the FODIBM for 1000 time steps.}
\label{table_cost}
\end{table}

\subsection{Supersonic flow past an adiabatic circular cylinder}
\label{sub_cyl1}

To further assess the accuracy and capability of the present FODIBM-R for predicting compressible flows, a series of validation cases is conducted. First, supersonic flow ($Ma_{\infty}=2$) past an adiabatic circular cylinder is considered. A bow shock forms in front of the cylinder. The computational setup is identical to that described in Section \ref{sub_recons}.

\begin{figure}
\centerline{\includegraphics[width=1.0\linewidth]{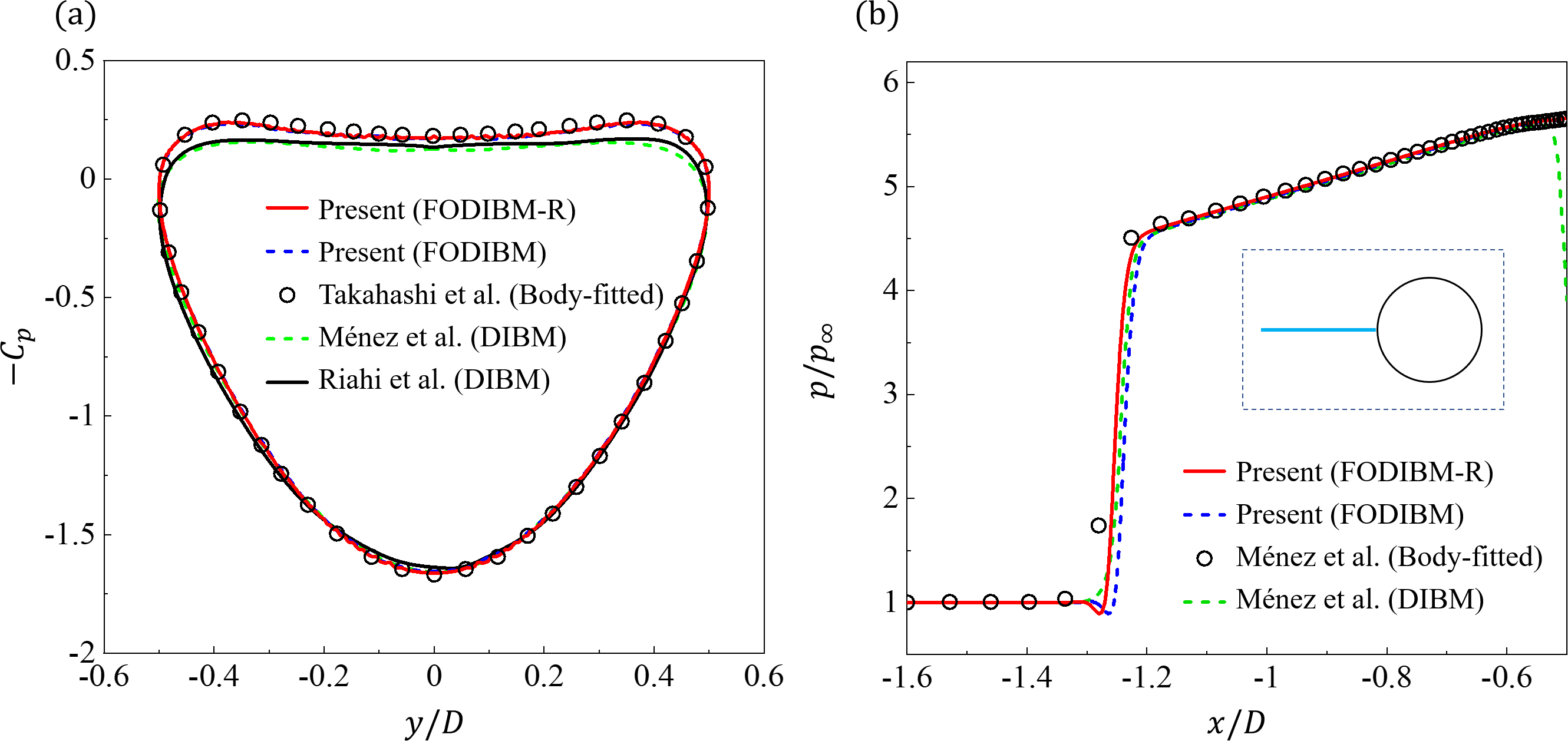}}
\caption{(a) Distribution of the pressure coefficient ($C_p$) on the cylinder surface and (b) pressure profile along the line $y/D = 0$ for supersonic flow past an adiabatic circular cylinder at $Ma_{\infty} = 2$. The stagnation point is located at $x/D = -0.5$. The data extraction line is indicated in blue (not to scale).}
\label{fig_cp_cylinder}
\end{figure}

\begin{figure}
\centerline{\includegraphics[width=1.0\linewidth]{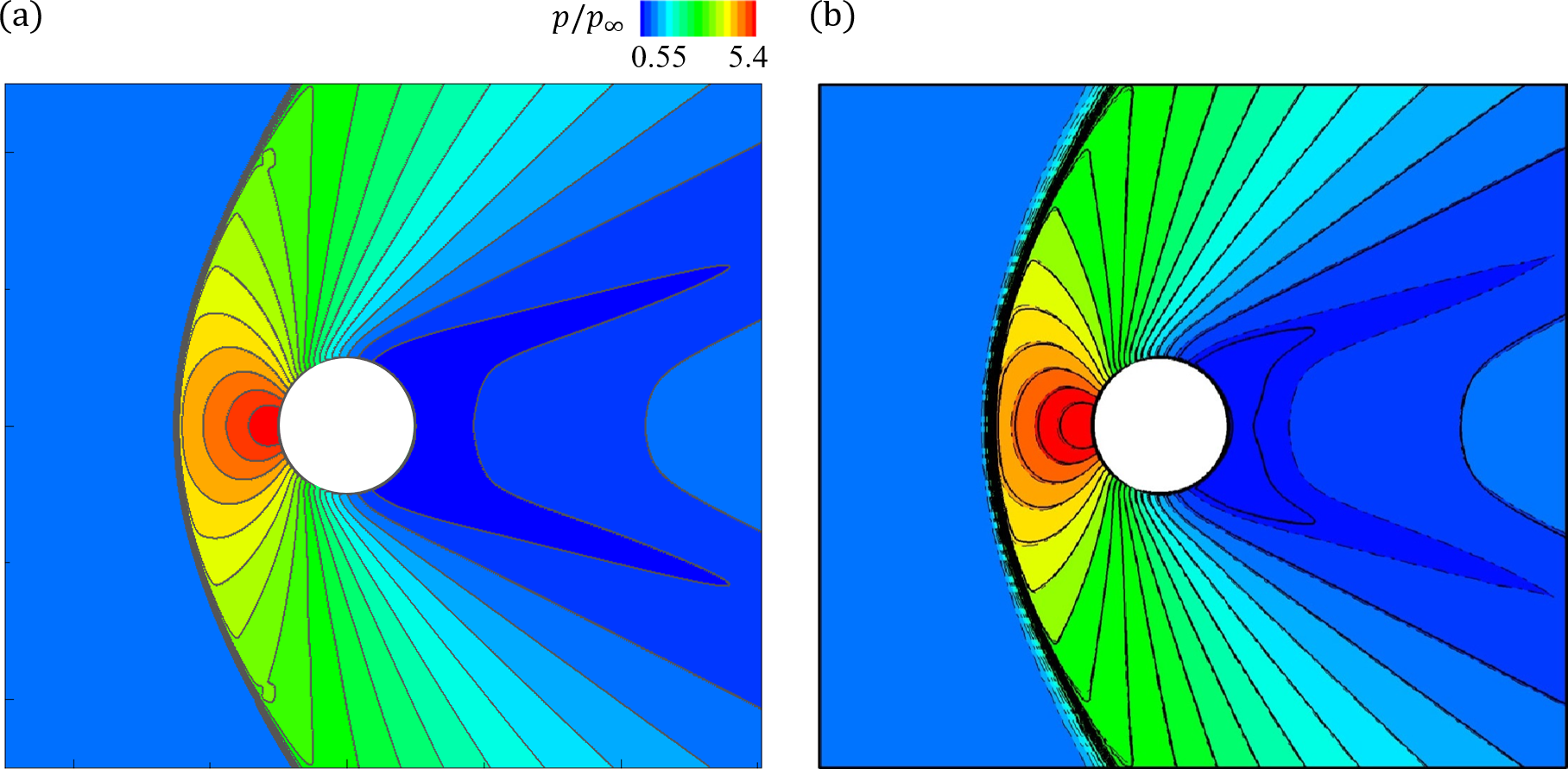}}
\caption{Pressure ($p$) contours. (a) Present results obtained using the FODIBM-R. (b) Background contours with dashed lines are generated by a body-fitted method \cite{yang2016hybrid}, while contours with solid lines are predicted by a DIBM \cite{sun2020diffuse}.}
\label{fig_contour_cylinder}
\end{figure}

The distribution of $C_p$ on the cylinder surface is shown in Fig. \ref{fig_cp_cylinder}(a). Reference results obtained using a body-fitted method and the DIBM are also included for comparison \cite{menez2023assessment,riahi2018pressure,takahashi2014numerical}. The results obtained by the FODIBM-R and the FODIBM are nearly identical. This is because the pressure variation near the cylinder surface is relatively small and thus insensitive to the slight shift of the effective boundary position. Both methods show good agreement with the body-fitted result \cite{takahashi2014numerical}. In contrast, the conventional DIBM \cite{menez2023assessment,riahi2018pressure} exhibits lower accuracy near the rear of the cylinder. To provide a more intuitive assessment of the accuracy of the FODIBM-R, pressure contour comparisons obtained using the FODIBM-R, a body-fitted method \cite{yang2016hybrid}, and the conventional DIBM \cite{sun2020diffuse} are presented in Fig. \ref{fig_contour_cylinder}. In Fig. \ref{fig_contour_cylinder}(b), the background contours with dashed lines correspond to the body-fitted method, while the solid lines represent the DIBM predictions. The results clearly show that the pressure field in the wake predicted by the DIBM exhibits significant deviation from the body-fitted solution. In contrast, the pressure field predicted by the FODIBM-R in Fig. \ref{fig_contour_cylinder}(a) agrees very well with the body-fitted results. This indicates that the boundary conditions are accurately enforced by the FODIBM-R.

The pressure distribution along the stagnation line ($y/D=0$) is presented in Fig. \ref{fig_cp_cylinder}(b). The FODIBM-R predicts the shock location and pressure jump ($y/D \approx -1.25$) more accurately than the FODIBM. This observation further confirms that the inaccurate prediction of the shock position originates from the inward shift of the effective boundary position. In addition, the conventional DIBM \cite{menez2023assessment} produces a noticeable pressure drop near the cylinder surface due to the diffusion effect, leading to reduced accuracy in Figs. \ref{fig_cp_cylinder}(a) and \ref{fig_contour_cylinder}. This issue is effectively eliminated in both the FODIBM-R and the FODIBM, as the forcing is applied only inside the boundary. Consequently, both methods yield results in good agreement with the body-fitted solution.

As shown in Fig. \ref{fig_cp_cylinder}(b), the oscillations observed near the shock ($y/D \approx -1.27$) are caused by the low artificial bulk viscosity used in the numerical setup. A detailed investigation of this issue has been reported in previous work \cite{mao2026explicit}. Although increasing the artificial bulk viscosity can suppress these oscillations, it also degrades the accuracy of shock capturing. This limitation is inherent to the numerical treatment and is not related to the IBM.

\begin{table}[t]
\centering
\small
\begin{tabular}{l c c}
  \hline
    & $C_d$ & $\Delta_s$ \\
  \hline
  Present (FODIBM-R) & 1.57 & 0.71 \\
  Present (FODIBM) & 1.55 & 0.70 \\
  Takahashi et al. (Body-fitted) & 1.55 & - \\
  Ménez et al. (Body-fitted) & 1.62 & 0.73 \\
  Ménez et al. (SIBM) & 1.59 & 0.73 \\
  Riahi et al. (DIBM) & 1.51 & 0.69 \\
  \hline
\end{tabular}
\caption{Drag coefficient ($C_d$) and shock stand-off distance ($\Delta_s$) for supersonic flow past an adiabatic circular cylinder.}
\label{table_cylinder}
\end{table}

To quantitatively assess the accuracy of the FODIBM-R, the drag coefficient $C_d$ and the shock stand-off distance $\Delta_s$ are reported in Table \ref{table_cylinder}. The shock stand-off distance $\Delta_s$ is defined as the streamwise distance between the stagnation point on the cylinder surface and the shock location, where the pressure ratio $p_s/p_{\infty}$ satisfies \cite{liepmann2001elements}

\begin{equation}
\frac{p_s}{p_{\infty}}=\frac{2\gamma}{\gamma+1}Ma_{\infty}^2-\frac{\gamma-1}{\gamma+1},
\end{equation}
where $\gamma=1.4$ is the specific heat ratio.

Compared with the FODIBM, both $C_d$ and $\Delta_s$ predicted by the FODIBM-R are higher. This is because the effective diameter of the cylinder is underestimated in the FODIBM, leading to an underestimation of the aerodynamic force. The present predictions of $C_d$ and $\Delta_s$ show good agreement with reference data obtained using body-fitted methods and the sharp-interface IBM (SIBM) \cite{menez2023assessment,takahashi2014numerical}, thereby demonstrating the effectiveness of the FODIBM-R in predicting compressible flows.

\subsection{Supersonic flow past an isothermal circular cylinder}
\label{sub_cyl2}

\begin{figure}
\centerline{\includegraphics[width=0.5\linewidth]{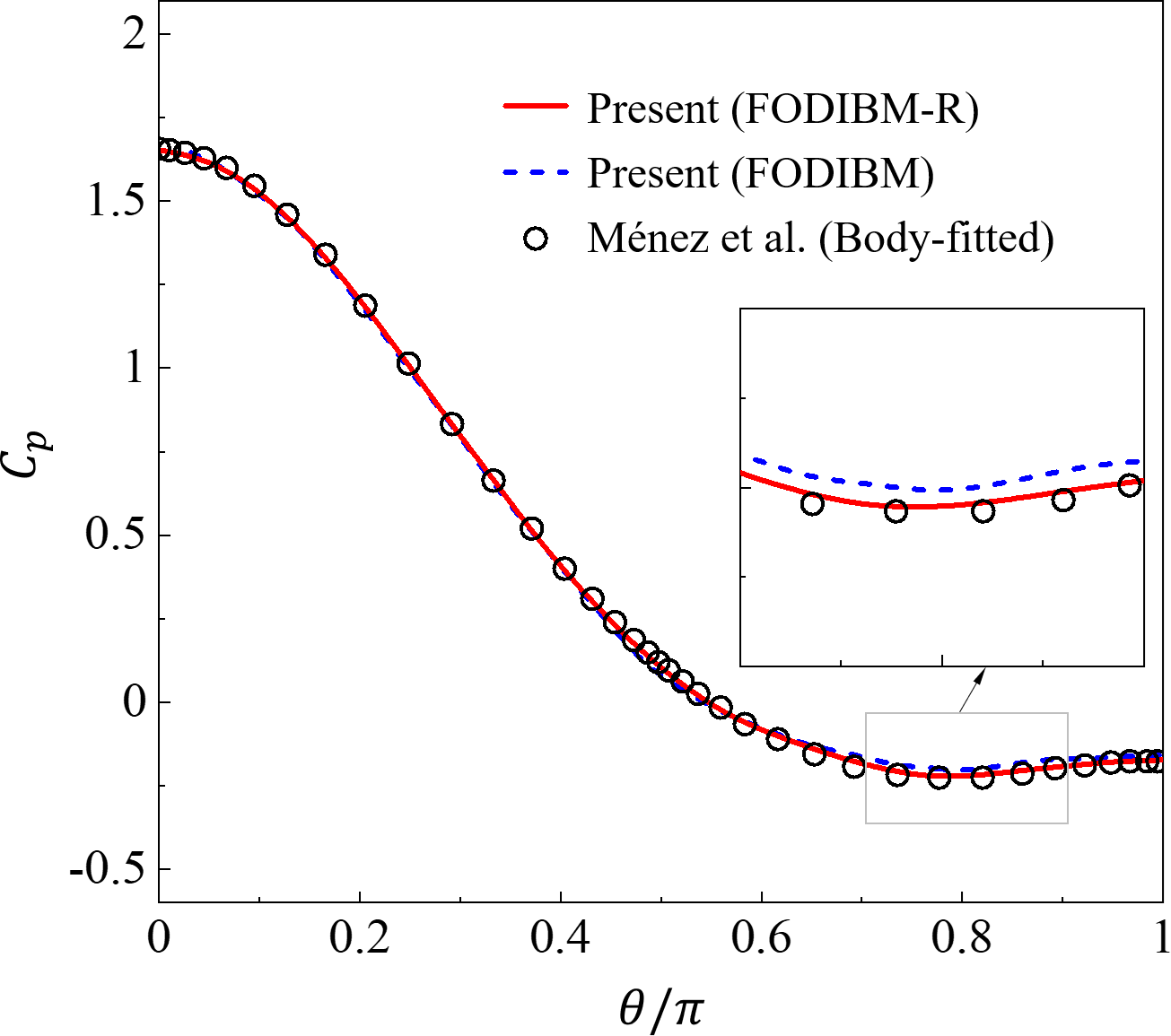}}
\caption{Distribution of the pressure coefficient ($C_p$) on the cylinder surface for supersonic flow past an isothermal circular cylinder at $Ma_{\infty}=2$ ($\theta/\pi=0$ and 1 at leading edge and trailing edge, respectively).}
\label{fig_cp_cylinder_iso}
\end{figure}

\begin{figure}
\centerline{\includegraphics[width=1.0\linewidth]{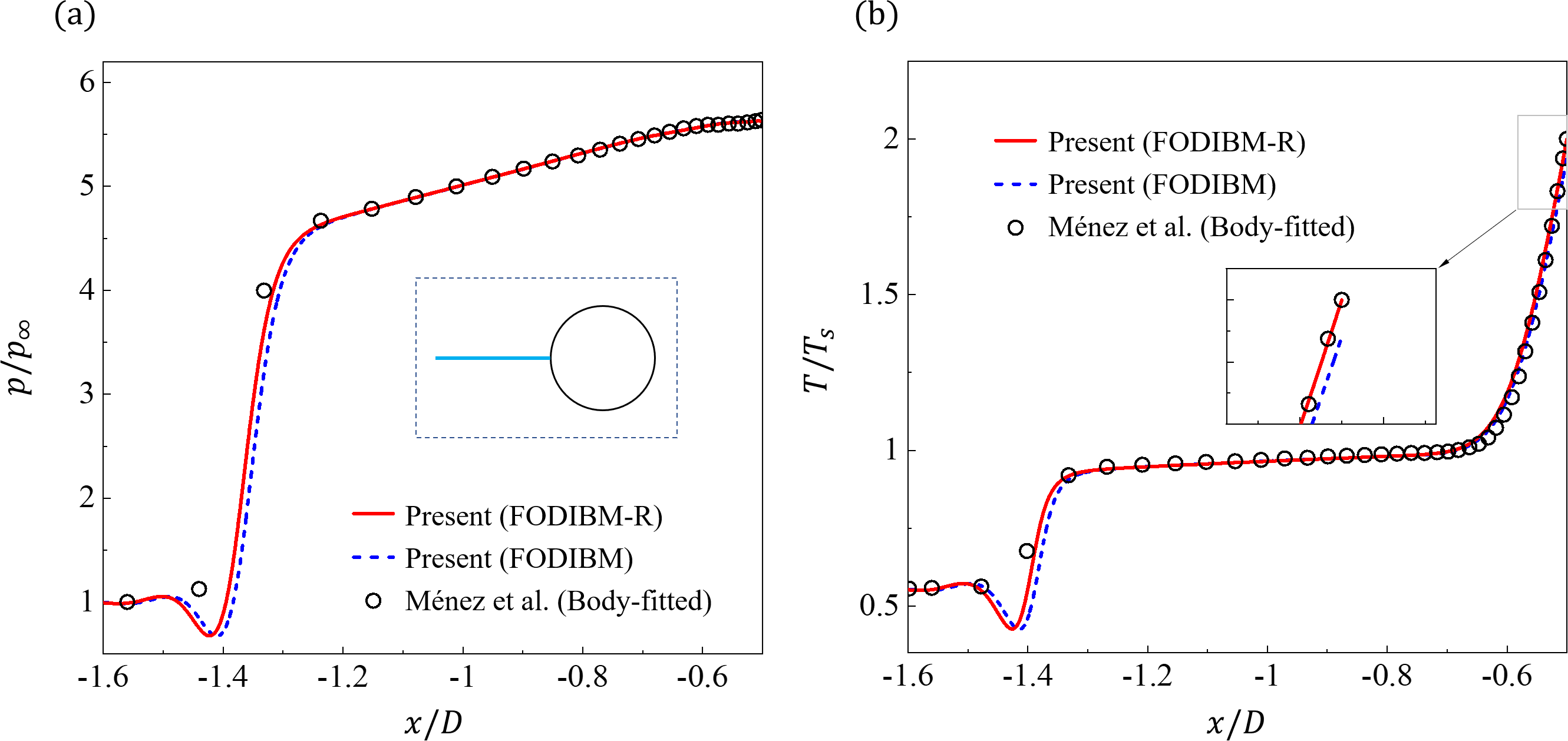}}
\caption{(a) Pressure and (b) temperature profiles along the line $y/D=0$ for supersonic flow past an isothermal circular cylinder at $Ma_{\infty}=2$. The stagnation point is located at $x/D=-0.5$. The data extraction line is indicated in blue (not to scale).}
\label{fig_pt_cylinder_iso}
\end{figure}

A supersonic flow over an isothermal circular cylinder is simulated to further evaluate the capability of the proposed method in enforcing thermal boundary conditions. The free-stream conditions are specified as $Ma_{\infty} = 2.0$ and $Re = 300$, with a temperature of $T_{\infty} = 162.8$ K, corresponding to a stagnation temperature of $T_s = 293$ K. On the cylinder surface, no-slip ($\boldsymbol{u}^t = 0$) and isothermal ($T^t = 2T_s$) boundary conditions are imposed. The cylinder is centered at $(0, 0)$ within a rectangular computational domain spanning $[-10D, 30D] \times [-10D, 10D]$. A locally refined mesh is employed, with a minimum grid spacing of $\Delta x_{\mathrm{min}} = D/200$ in the subregion $[-2D, 2D] \times [-2D, 2D]$.

Figure \ref{fig_cp_cylinder_iso} shows the distribution of the pressure coefficient $C_p$. The prediction of $C_p$ near the trailing edge of the cylinder ($0.6 < \theta/\pi < 1$) is improved by the FODIBM-R compared with the FODIBM, which is attributed to the more accurate enforcement of the boundary conditions. The FODIBM-R results agree well with the reference data \cite{menez2023assessment}. Figure \ref{fig_pt_cylinder_iso} presents the pressure and temperature profiles along the stagnation line ($y/D=0$). Overall, the results obtained by the FODIBM-R are similar to those of the FODIBM. However, the FODIBM-R provides a slightly improved prediction of the shock. In particular, the temperature near the wall is more accurately resolved.

To quantitatively assess the accuracy of the FODIBM-R, the isothermal boundary condition error is defined as
\begin{equation}
\epsilon_{iso} = \frac{|T^s - T^t|}{T^t}.
\end{equation}

Compared with the FODIBM ($\bar{\epsilon}_{iso} = 1.2 \times 10^{-2}$), the FODIBM-R reduces the space-averaged isothermal boundary condition error to $\bar{\epsilon}_{iso} = 1.8 \times 10^{-3}$, corresponding to a reduction of $85\%$. This level of improvement is comparable to that observed for the no-slip boundary condition error in Section \ref{sub_bcerr}. These results further confirm the accuracy and suitability of the proposed FODIBM-R for simulating compressible flows.

\subsection{Supersonic flow past adiabatic 5-point stars}
\label{sub_star}

\begin{figure}
\centerline{\includegraphics[width=1.0\linewidth]{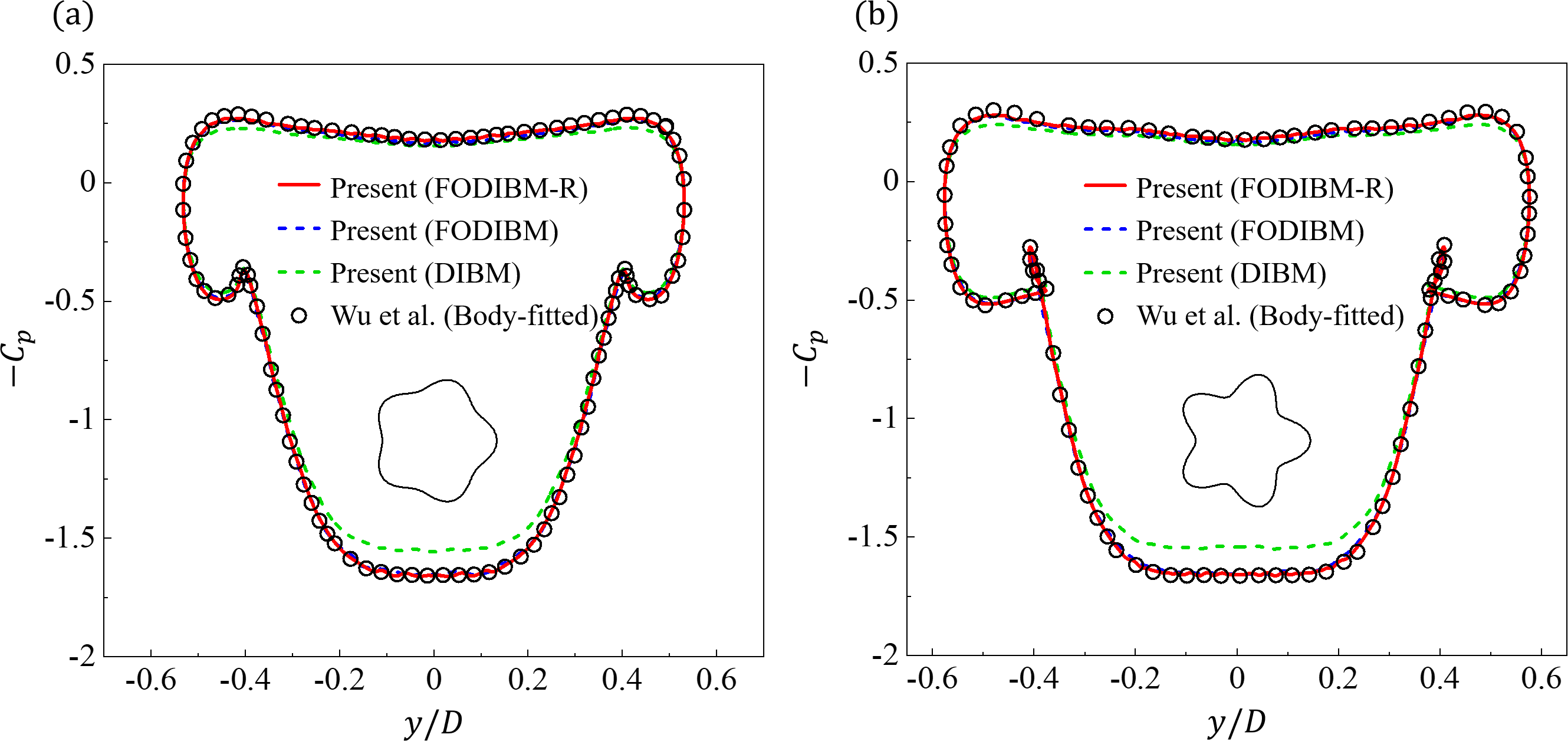}}
\caption{Distribution of the pressure coefficient ($C_p$) on the surface of 5-point stars at (a) $\Delta r/D=0.05$ and (b) $\Delta r/D=0.1$.}
\label{fig_cp_star}
\end{figure}

To evaluate the performance of the method for more complex geometries, supersonic flows over 5-point star-shaped bodies are simulated. The geometry of each star is defined by
\begin{equation}
\binom{X_l}{Y_l}=\binom{\left[r_0+\Delta r\mathrm{cos}(5\theta)\right]\mathrm{cos}(\theta)}{\left[r_0+\Delta r\mathrm{cos}(5\theta)\right]\mathrm{sin}(\theta)},
\end{equation}
where $r_0$ denotes the mean radius and $\Delta r$ represents the amplitude of the radial perturbation, with the reference diameter defined as $D = 2r_0$. Two configurations are considered, corresponding to $\Delta r/D = 0.05$ and $0.1$. The free-stream Mach and Reynolds numbers, based on $D$, are specified as $Ma_\infty = 2.0$ and $Re = 200$. No-slip ($\boldsymbol{u}^t = 0$) and adiabatic ($\partial T / \partial n = 0$) boundary conditions are applied on the star surfaces. Each star is centered at $(0,0)$ within a rectangular computational domain $[-10D, 30D] \times [-10D, 10D]$, with a locally refined mesh of $\Delta x_{\mathrm{min}} = D/100$ in the subregion $[-2D, 2D] \times [-2D, 2D]$.

Fig. \ref{fig_cp_star} presents the distribution of $C_p$ on the surfaces of 5-point stars for $\Delta r/D = 0.05$ and $0.1$, with results from the body-fitted method \cite{wu2025one} included for reference. In both cases, the FODIBM-R predictions show good agreement with the body-fitted results. Notably, FODIBM-R provides slight improvements over FODIBM near the trailing edge, and substantial improvements over the conventional DIBM near both the leading and trailing edges.

\begin{figure}
\centerline{\includegraphics[width=1.0\linewidth]{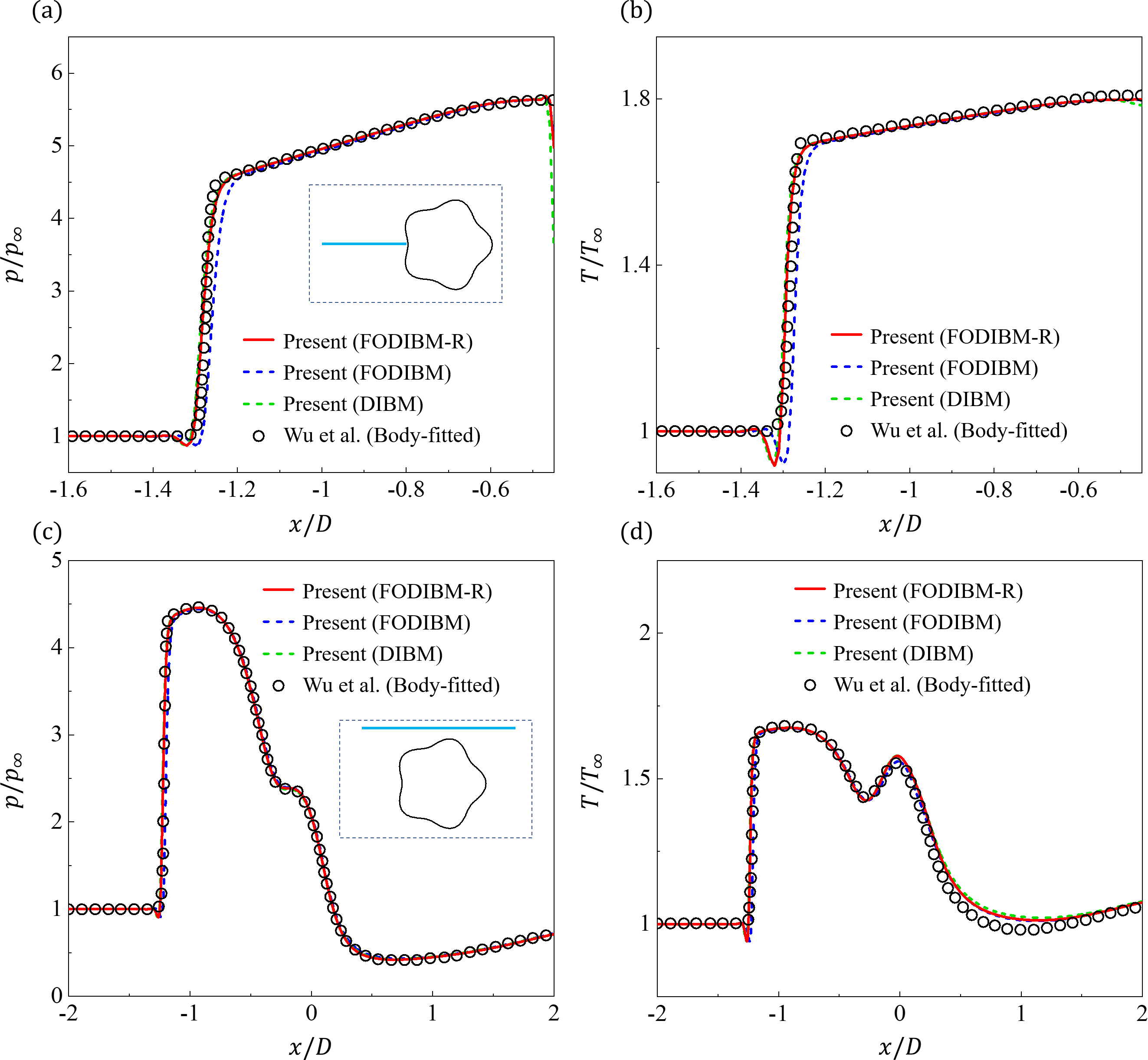}}
\caption{(a) Pressure and (b) temperature profile along the line $y/D=0$. (c) Pressure and (d) temperature profile along the line $y/D=0.6$ for supersonic flow past an adiabatic 5-point star at $\Delta r/D=0.05$. The stagnation point is located at $x/D=-0.45$. The data extraction line is indicated in blue (not to scale).}
\label{fig_pt_star1}
\end{figure}

\begin{figure}
\centerline{\includegraphics[width=1.0\linewidth]{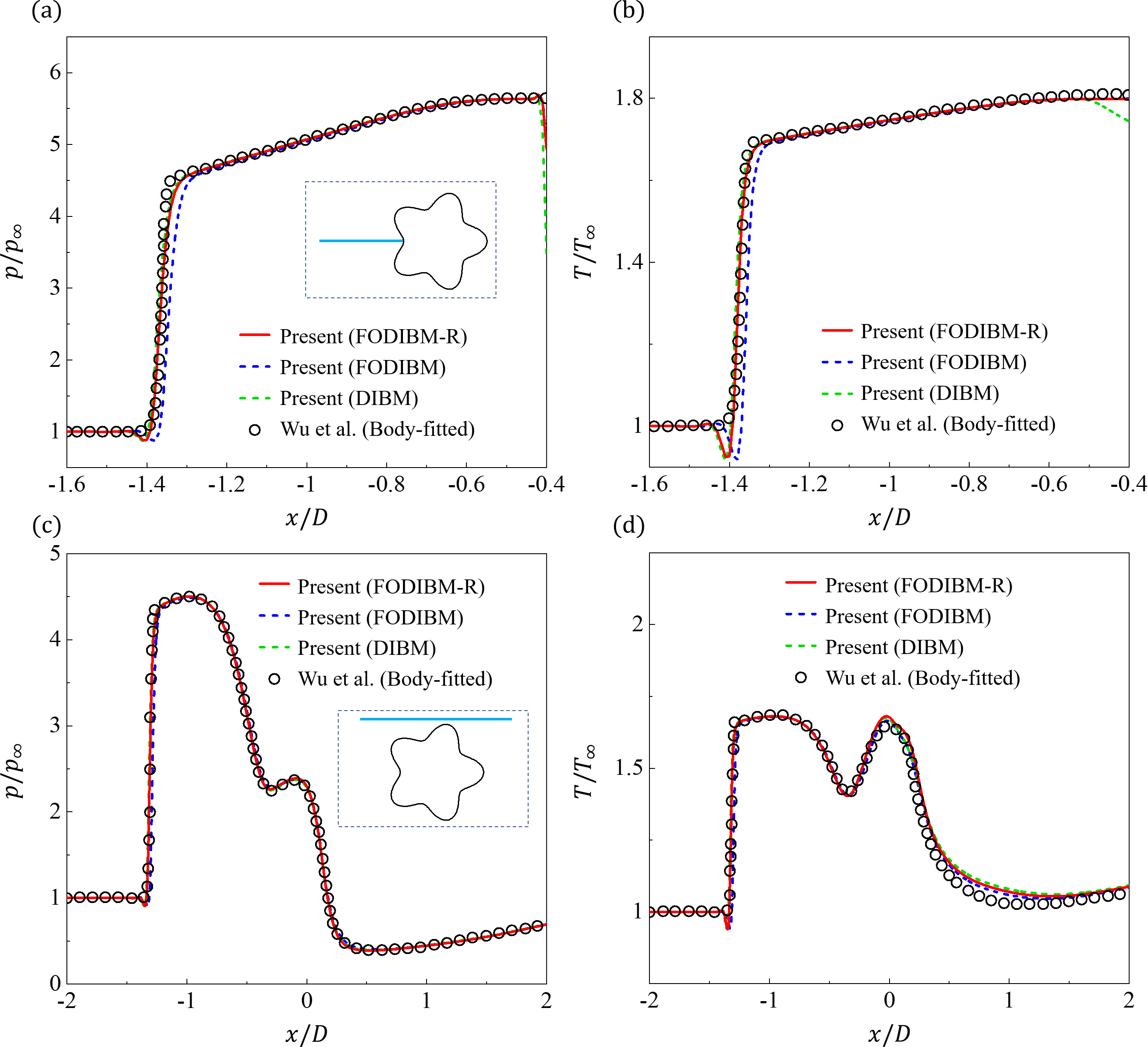}}
\caption{(a) Pressure and (b) temperature profile along the line $y/D=0$. (c) Pressure and (d) temperature profile along the line $y/D=0.6$ for supersonic flow past an adiabatic 5-point star at $\Delta r/D=0.1$. The stagnation point is located at $x/D=-0.4$. The data extraction line is indicated in blue (not to scale).}
\label{fig_pt_star2}
\end{figure}

Figs. \ref{fig_pt_star1} and \ref{fig_pt_star2} present the pressure and temperature profiles for $\Delta r/D=0.05$ and $\Delta r/D=0.1$. Overall, the FODIBM-R predictions outperform both FODIBM and DIBM and show good agreement with the body-fitted results \cite{wu2025one}. Specifically, FODIBM-R captures the shock position and the associated pressure and temperature jumps more accurately than FODIBM, while its near-wall predictions of pressure and temperature are comparable to FODIBM and more accurate than DIBM. The remaining pressure drop near the surface is attributed to the star's complex geometry and the absence of a zero-pressure-gradient condition at the geometric boundary. Notably, DIBM fails to enforce the adiabatic boundary condition, resulting in a spurious temperature drop near the star surface (Figs. \ref{fig_pt_star1}(b) and \ref{fig_pt_star2}(b)), whereas both FODIBM-R and FODIBM correctly impose the adiabatic condition. The slight deviations in the wake temperature from the body-fitted results (Figs. \ref{fig_pt_star1}(d) and \ref{fig_pt_star2}(d)) are likely due to differences in mesh configuration. These results demonstrate that FODIBM-R effectively combines the advantages of FODIBM and DIBM, namely low diffusion and accurate enforcement of the effective boundary position, which highlights its capability for simulating compressible flows around complex geometries.

To quantitatively evaluate the improvement, Table \ref{table_star} summarizes the shock stand-off distance ($\Delta_s$) and the relative errors compared with the body-fitted method. Compared with FODIBM, FODIBM-R reduces the relative errors by more than $50\%$ for both configurations ($\Delta r/D = 0.05$ and $0.1$), confirming its enhanced accuracy. Although the absolute error appears moderate, such a reduction is significant for shock-position prediction, since shocks are intrinsically sharp discontinuities with negligible physical thickness and are highly sensitive to boundary representation.

\begin{table}[t]
\centering
\small
\begin{tabular}{l c c c c}
  \hline
   & $\Delta_s$ ($\Delta r=0.05$) & $err$ $(\%)$ & $\Delta_s$ ($\Delta r=0.1$) & $err$ $(\%)$ \\
  \hline
  Present (FODIBM-R) & 0.78 & 1.3 & 0.91 & 2.1 \\
  Present (FODIBM) & 0.76 & 3.7 & 0.89 & 4.2 \\
  Wu et al. (Body-fitted) & 0.79 & - & 0.93 & - \\
  \hline
\end{tabular}
\caption{Shock stand-off distance ($\Delta_s$) for supersonic flow past an adiabatic 5-point star.}
\label{table_star}
\end{table}

\subsection{Compressible flows past a NACA0012 airfoil}
\label{sub_naca0012_compres}

\begin{figure}
\centerline{\includegraphics[width=1.0\linewidth]{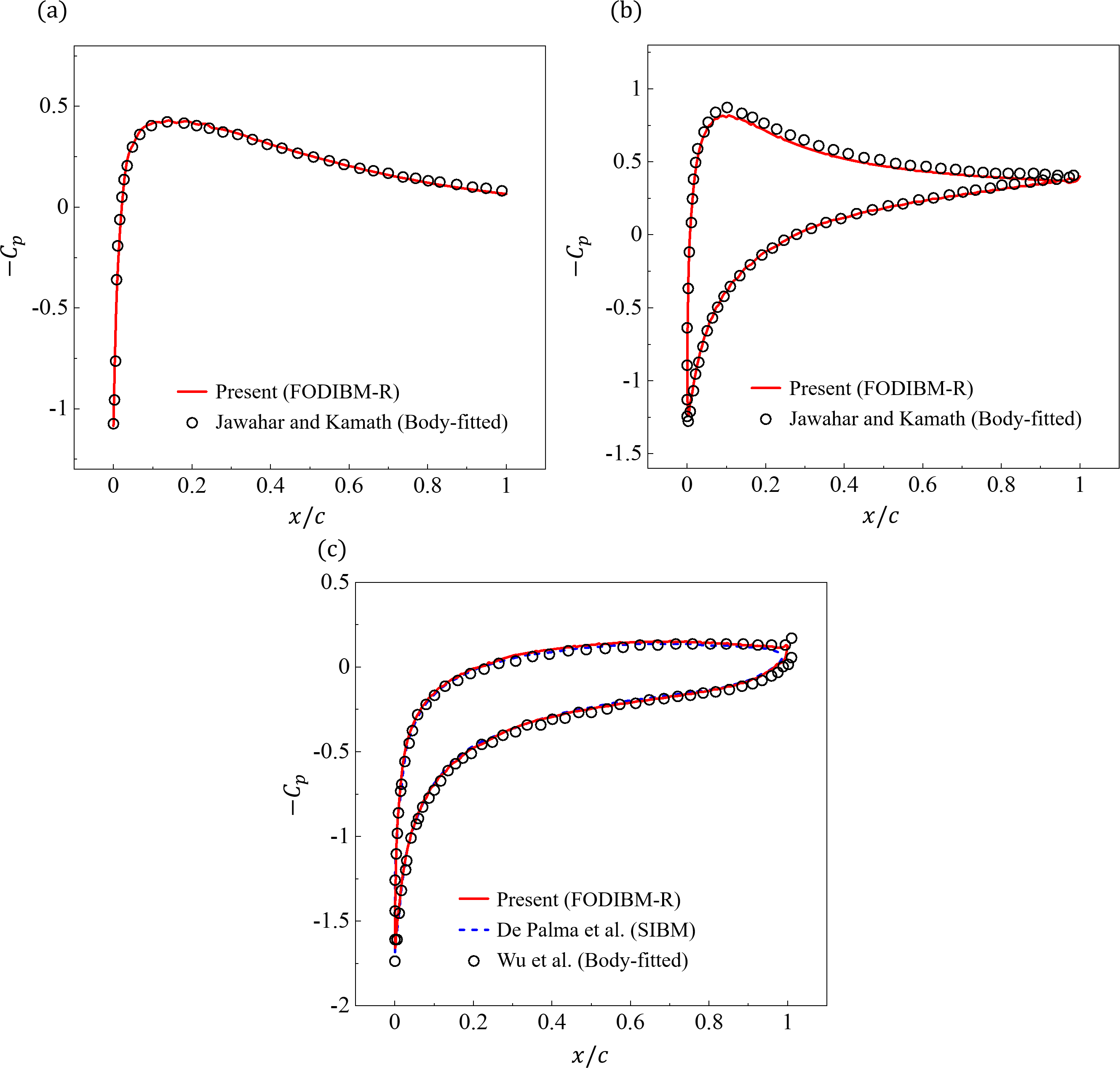}}
\caption{Distribution of the pressure coefficient ($C_p$) on the surface of a NACA0012 airfoil at (a) $Ma = 0.5$, $Re = 5000$, $AoA = 0^\circ$; (b) $Ma = 0.8$, $Re = 500$, $AoA = 10^\circ$; and (c) $Ma = 2.0$, $Re = 1000$, $AoA = 10^\circ$.}
\label{fig_cp_naca0012}
\end{figure}

\begin{figure}
\centerline{\includegraphics[width=1.0\linewidth]{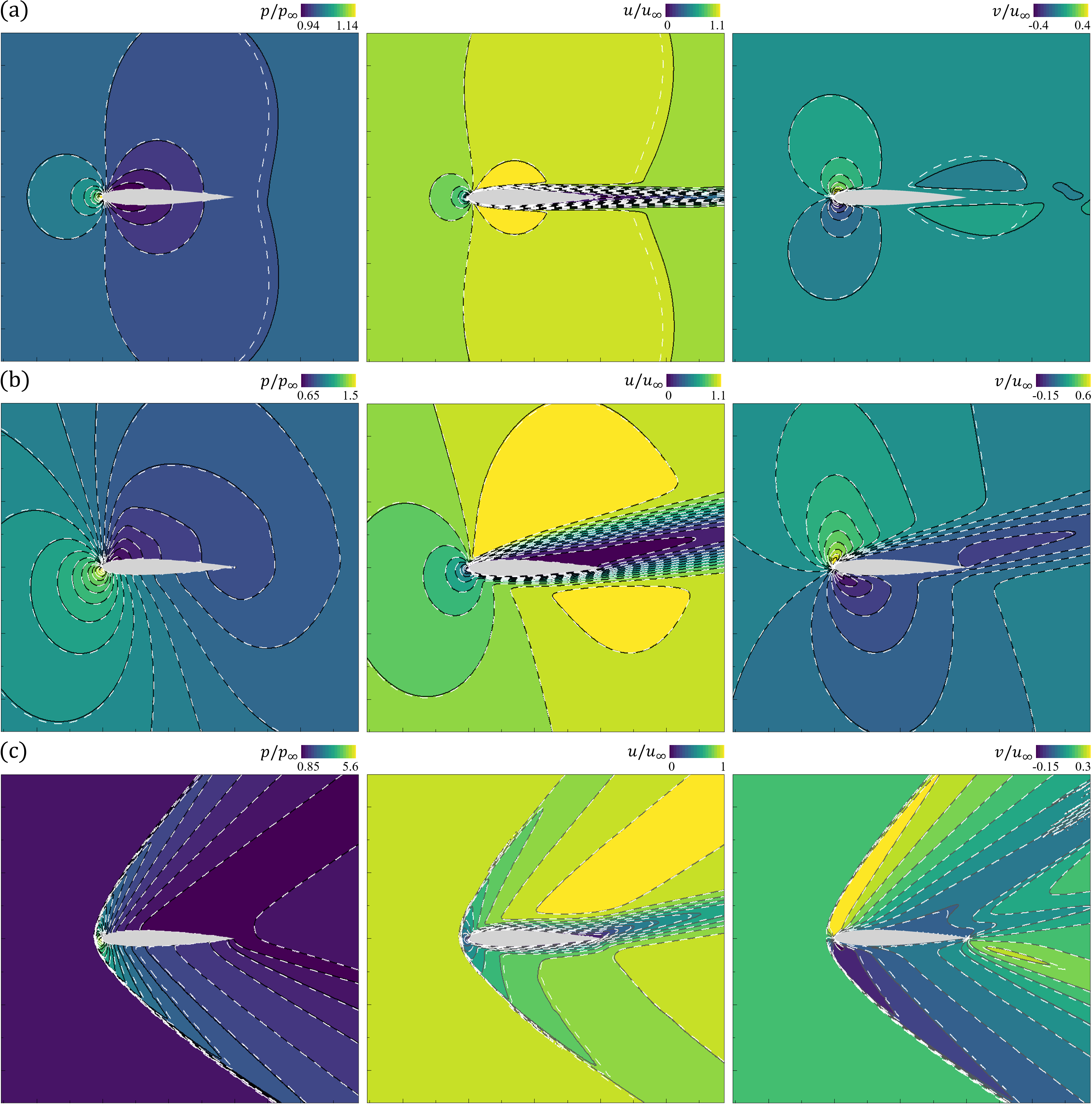}}
\caption{Pressure ($p$), streamwise velocity ($u$), and spanwise velocity ($v$) contours of compressible flow past a NACA0012 airfoil: (a) $Ma = 0.5$, $Re = 5000$, $AoA = 0^\circ$, (b) $Ma = 0.8$, $Re = 500$, $AoA = 10^\circ$, and (c) $Ma = 2$, $Re = 1000$, $AoA = 10^\circ$. The background contours with black solid lines correspond to the hybrid LBM results, while the contours with white dashed lines represent the FODIBM-R predictions.}
\label{fig_naca0012_contour}
\end{figure}

In this part, the FODIBM-R is further validated using compressible flows past a NACA0012 airfoil. Three configurations, ranging from subsonic to supersonic regimes, are considered: (a) $Ma_\infty = 0.5$, $Re = \rho_{\infty} u_{\infty} c / \mu = 5000$, and $AoA = 0^\circ$; (b) $Ma_\infty = 0.8$, $Re = 500$, and $AoA = 10^\circ$; (c) $Ma_\infty = 2.0$, $Re = 1000$, and $AoA = 10^\circ$, where $c$ and $AoA$ denote the chord length and the angle of attack, respectively. No-slip ($\boldsymbol{u}^t = 0$) and adiabatic ($\partial T / \partial n = 0$) boundary conditions are applied on the airfoil surface. The leading edge of the NACA0012 airfoil is located at $(0,0)$ within a rectangular computational domain $[-10D, 30D] \times [-10D, 10D]$, with a locally refined mesh of $\Delta x_{\mathrm{min}} = D/256$ in the subregion $[-0.5D, 1.5D] \times [-D, D]$. The angle of attack is varied by adjusting the inflow direction.

Figure \ref{fig_cp_naca0012} presents the distribution of the pressure coefficient ($C_p$) on the surface of the NACA0012 airfoil for the three configurations. The present results show good agreement with those obtained using body-fitted methods and the SIBM \cite{wu2025one,jawahar2000high,de2006immersed}, confirming the accuracy of the FODIBM-R in predicting the near-wall flow field. To further assess the global accuracy, comparisons of pressure, streamwise velocity, and spanwise velocity contours obtained using the FODIBM-R and the hybrid LBM without IBM are presented in Fig. \ref{fig_naca0012_contour}. The accuracy of the hybrid LBM has been validated in various previous studies \cite{wissocq2022restoring,feng2019solid,tsetoglou2025mass}. The background contours with black solid lines correspond to the hybrid LBM results, while the contours with white dashed lines represent the FODIBM-R predictions. Overall, the FODIBM-R results show good agreement with the reference data obtained using the hybrid LBM. In particular, the FODIBM-R accurately captures the recirculation bubbles in the subsonic cases and the shock position in the supersonic case, demonstrating its capability for predicting the global flow field. It should be noted that the discrepancy observed in the wake region of Fig. \ref{fig_naca0012_contour}(a) is related to the discretization in the hybrid LBM. The trailing edge of the airfoil has a thickness of $2\Delta x$, which leads to weak vortex shedding, as confirmed by the spanwise velocity contour in Fig. \ref{fig_naca0012_contour}(a). In contrast, the FODIBM-R yields a sharper trailing edge and accurately reproduces the steady flow.

\begin{table}[t]
\centering
\small
\begin{tabular}{l c c}
  \hline
    & $C_d$ & $C_l$ \\
  \hline
  $Ma = 0.5$, $Re = 5000$, $AoA = 0^\circ$ &   &   \\
  Present (FODIBM-R) & 0.0533 & 0 \\
  Jawahar and Kamath (Body-fitted) & 0.05557 & 0 \\
  Mavriplis and Jameson (Body-fitted) & 0.0561 & 0 \\
  $Ma = 0.8$, $Re = 500$, $AoA = 10^\circ$ &   &   \\
  Present (FODIBM-R) & 0.2638 & 0.4196 \\
  Bristeau et al. (Body-fitted) & 0.243-0.2868 & 0.4145-0.5170 \\
  Jawahar and Kamath (Body-fitted) & 0.27216 & 0.49394 \\
  $Ma = 2.0$, $Re = 1000$, $AoA = 10^\circ$ &   &   \\
  Present (FODIBM-R) & 0.2501 & 0.3431 \\
  Bristeau et al. (Body-fitted) & 0.2515-0.2535 & 0.3388-0.3427 \\
  Forsyth and Jiang (Body-fitted) & 0.2548 & 0.3407 \\
  \hline
\end{tabular}
\caption{Drag coefficient ($C_d$) and lift coefficient ($C_l$) for compressible flow past a NACA0012 airfoil.}
\label{table_naca0012}
\end{table}

To quantitatively evaluate the accuracy of the FODIBM-R in predicting aerodynamic forces, the drag and lift coefficients are summarized in Table \ref{table_naca0012}. Reference results obtained using body-fitted methods \cite{jawahar2000high,mavriplis1990multigrid,bristeau2013numerical,forsyth1997nonlinear} are included for comparison. For all three configurations, the present results show good agreement with the reference data, demonstrating the reliability of the FODIBM-R in predicting aerodynamic forces across a wide range of flow regimes.

\subsection{Supersonic flow past a three-dimensional sphere}

\begin{figure}
\centerline{\includegraphics[width=1.0\linewidth]{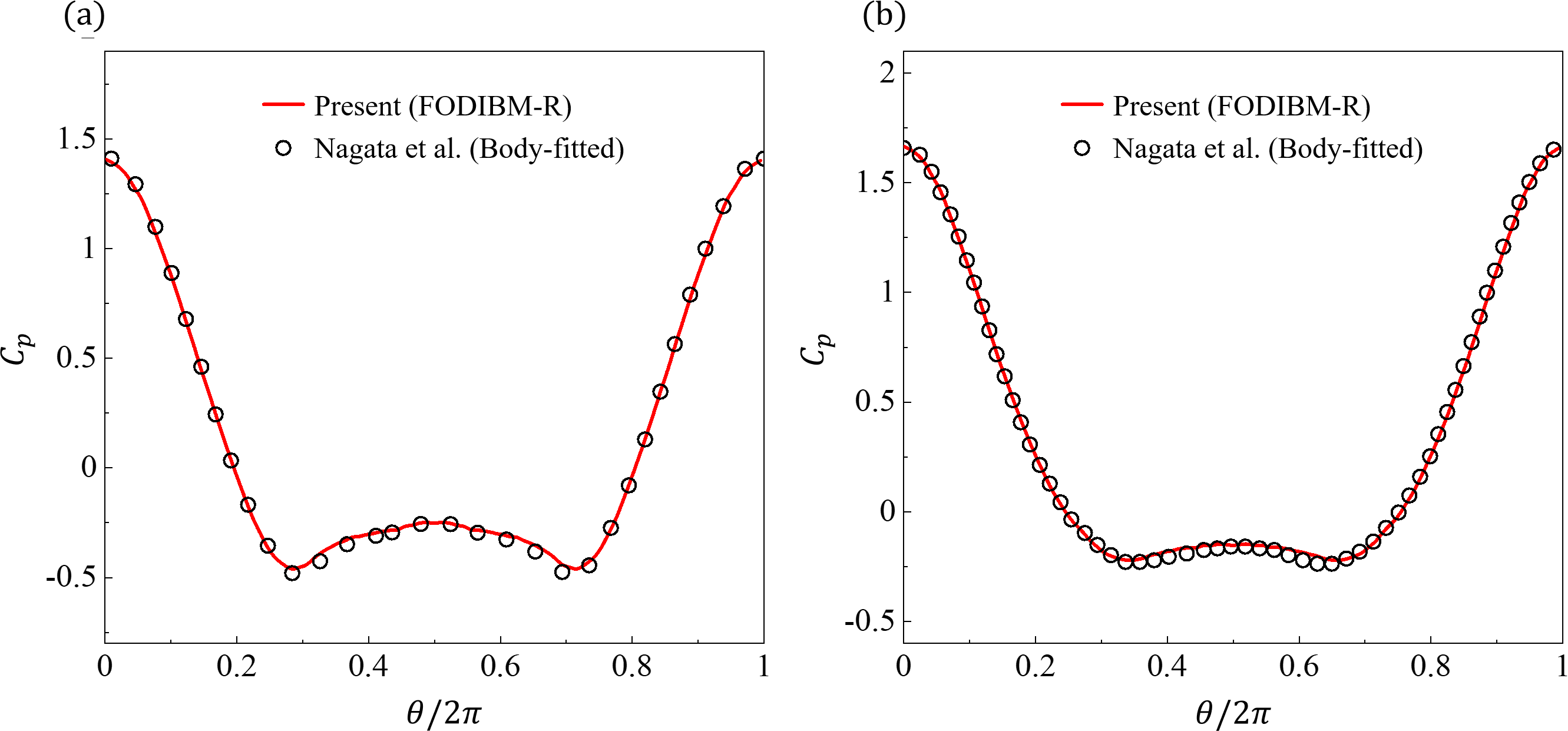}}
\caption{Distribution of the pressure coefficient ($C_p$) on the surface of a sphere above the equator at (a) $Ma_{\infty} = 1.2$ and (b) $Ma_{\infty} = 2$ ($\theta/2\pi = 0$ and $1$ correspond to the leading edge).}
\label{fig_sphere_ma1.2_2}
\end{figure}

\begin{table}[t]
\centering
\small
\begin{tabular}{l c c}
  \hline
    & $C_d$ & $\Delta_s$ \\
  \hline
  $Ma_{\infty} = 1.2$ & & \\
  Present (FODIBM-R) & 1.287 & 0.678 \\
  Nagata et al. (Body-fitted) & 1.283 & 0.690 \\
  Heberle et al. (Exp) & - & 0.700 \\
  $Ma_{\infty} = 2.0$ & & \\
  Present (FODIBM-R) & 1.378 & 0.192 \\
  Nagata et al. (Body-fitted) & 1.382 & 0.200 \\
  \hline
\end{tabular}
\caption{Drag coefficient ($C_d$) and shock stand-off distance ($\Delta_s$) for supersonic flow past a sphere.}
\label{table_sphere_ma1.2_2}
\end{table}

The performance of the FODIBM-R in three-dimensional configurations is further assessed through simulations of supersonic flow over a sphere. The sphere with diameter $D$ is centered at $(0,0,0)$ within a rectangular computational domain spanning $[-10D, 30D] \times [-10D, 10D] \times [-10D, 10D]$. No-slip and adiabatic boundary conditions are applied on the solid surface. A locally refined mesh is adopted, with a minimum grid spacing of $\Delta x_{\mathrm{min}} = D/64$ in the region $[-1.5D, 1.5D] \times [-1.5D, 1.5D]\times [-1.5D, 1.5D]$. The Reynolds number is set to $Re = 300$, and two free-stream Mach numbers, $Ma_{\infty} = 1.2$ and $2.0$, are considered. Figure \ref{fig_sphere_ma1.2_2} shows the distribution of the pressure coefficient ($C_p$) on the sphere surface above the equator, demonstrating good agreement with reference data \cite{nagata2018direct}. Moreover, the drag coefficient $C_d$ and the shock stand-off distance $\Delta_s$ are summarized in Table \ref{table_sphere_ma1.2_2}. Both quantities closely match the results obtained from body-fitted simulations and experimental data \cite{nagata2016investigation,heberle1949data}. This confirms the accuracy and robustness of the FODIBM-R for three-dimensional supersonic flows.

\subsection{Subsonic flow past a rotating sphere}

\begin{figure}
\centerline{\includegraphics[width=1.0\linewidth]{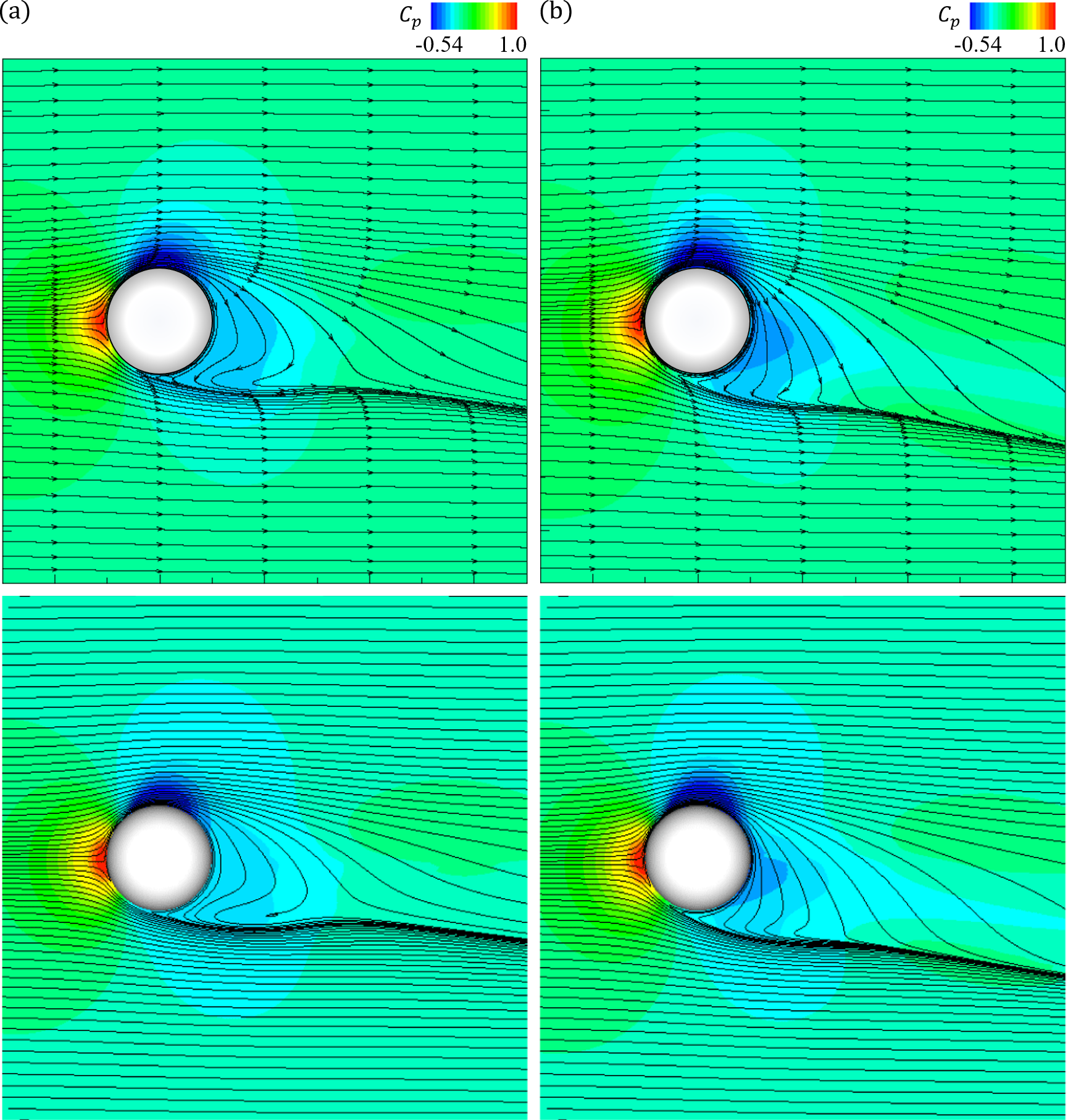}}
\caption{Pressure coefficient ($C_p$) distributions and streamlines of the time-averaged flow field for subsonic flow past a rotating sphere at (a) $\Omega^* = 0.3$ and (b) $\Omega^* = 0.6$. First row: present results obtained using the FODIBM-R. Second row: reference results generated by a body-fitted method \cite{nagata2018direct}.}
\label{fig_rotsphere}
\end{figure}

\begin{table}[t]
\centering
\small
\begin{tabular}{l c c c}
  \hline
    & $C_d$ & $C_l$ & $St$ \\
  \hline
  $\Omega^* = 0.3$ &   &   &   \\
  Present (FODIBM-R) & 0.750 & 0.332 & 0.202 \\
  Giacobello et al. (Body-fitted) & 0.744 & 0.345 & 0.215 \\
  Nagata et al. (Body-fitted) & 0.750 & 0.287 & 0.200 \\
  $\Omega^* = 0.6$ &   &   &   \\
  Present (FODIBM-R) & 0.862 & 0.512 & - \\
  Giacobello et al. (Body-fitted) & 0.850 & 0.515 & - \\
  Nagata et al. (Body-fitted) & 0.828 & 0.471 & - \\
  \hline
\end{tabular}
\caption{Drag coefficient ($C_d$), lift coefficient ($C_l$), and Strouhal number ($St$) for subsonic flow past a rotating sphere.}
\label{table_rotphere}
\end{table}

In this section, subsonic flow past a rotating sphere is further simulated to evaluate the accuracy of the FODIBM-R. The sphere with diameter $D$ is centered at $(0,0,0)$ within a rectangular computational domain spanning $[-20D, 60D] \times [-20D, 20D] \times [-20D, 20D]$. No-slip and adiabatic boundary conditions are applied on the sphere surface. A locally refined mesh is employed, with a minimum grid spacing of $\Delta x_{\mathrm{min}} = D/64$ in the region $[-D, 3D] \times [-D, D] \times [-D, D]$. The free-stream Mach and Reynolds numbers are specified as $Ma_\infty = 0.3$ and $Re = 300$. The sphere rotates with constant angular velocity $\omega$ around the $z$-axis. The dimensionless velocity ratio, $\Omega^* = u/u_{\infty} = \omega D / (2 u_{\infty})$, denotes the ratio of the surface velocity above the equator to the free-stream velocity. Two velocity ratios, $\Omega^* = 0.3$ and $0.6$, are considered, corresponding to cases with and without vortex shedding, respectively.

Fig. \ref{fig_rotsphere} presents the time-averaged pressure coefficient ($C_p$) distributions and streamlines for $\Omega^* = 0.3$ and $0.6$. The first row shows the present FODIBM-R results, while the second row shows reference results obtained using a body-fitted method \cite{nagata2018direct}. The present results agree well with the reference data. In particular, the FODIBM-R accurately captures the asymmetric wake, as well as the near-wall and far-field pressure distributions, demonstrating that boundary conditions are correctly enforced. Table \ref{table_rotphere} summarizes the drag coefficient ($C_d$), lift coefficient ($C_l$), and Strouhal number ($St$). Overall, the present results are in good agreement with reference data \cite{nagata2018direct,giacobello2009wake}. The FODIBM-R accurately predicts the aerodynamic forces and reproduces vortex shedding, further confirming its accuracy and capability for unsteady flow simulations.

\section{Conclusions}
\label{conclu}

In this study, a target-value reconstruction strategy is developed for the fully one-sided diffuse-interface immersed boundary method (FODIBM) to improve simulations of compressible flows around complex geometries. Implemented within a hybrid lattice Boltzmann framework, the strategy is applicable to both Dirichlet and Neumann boundary conditions.

A detailed analysis shows that the asymmetric support of the one-sided spreading operator causes the effective boundary to shift inward relative to the geometric boundary. To compensate for this displacement, the target values imposed at the effective boundary are reconstructed and subsequently used to update the corresponding immersed-boundary forcing terms, while the original interpolation and spreading operations are retained. Despite being confined to the target-value evaluation step, the proposed modification consistently reduces both the $\mathrm{L}_2$ and $\mathrm{L}_{\infty}$ error norms relative to the original FODIBM while preserving approximately second-order grid convergence. In particular, the no-slip and isothermal boundary-condition errors are reduced by $77\%$ and $85\%$, respectively. These improvements are achieved with negligible additional computational cost, while retaining the geometric flexibility and implementation simplicity of diffuse-interface IBMs.

The proposed reconstruction is assessed using a series of two- and three-dimensional compressible-flow cases involving a range of geometries and boundary conditions. Compared with the conventional diffuse-interface IBM and the original FODIBM, the improved method provides more accurate predictions of near-wall quantities and shock locations. The numerical results agree well with body-fitted reference solutions and experimental data. Overall, the proposed target-value reconstruction substantially improves accuracy at negligible additional computational cost, thereby enabling accurate and efficient simulations of complex compressible-flow configurations that remained challenging for the original FODIBM.

Future work will extend the present approach to more realistic aeronautical configurations and flow conditions.

\vspace{2ex} \noindent \textbf{Declaration of competing interest}

\vspace{2ex} The authors declare that they have no known competing financial interests or personal relationships that could have appeared to influence the work reported in this paper.

\vspace{2ex} \noindent \textbf{Acknowledgements}

\vspace{2ex} This research was supported by the FALCON project funded by the European Union's Horizon Europe research and innovation programme under grant agreement No 101138305, and by the ANR, Airbus, Fives-Pillard and SafranTech through the Industrial Chair Program LIBERTY ANR-23-CHIN-0005. Centre de Calcul Intensif d'Aix-Marseille and GENCI-TGCC/CINES (Grant A0152B11951) are acknowledged for granting access to their high-performance computing resources.

\vspace{2ex} \noindent \textbf{Data availability}

\vspace{2ex} Data will be made available on request.

\appendix
\section{D3Q19 Basis}
\label{app_d3q19}

The discrete velocities $\boldsymbol{c}_i$ of the $D3Q19$ scheme are defined as
\begin{equation}
\begin{gathered}
  c_{i,x}=\{0,1,1,0,-1,0,0,0,1,-1,-1,-1,0,1,0,0,0,-1,1\},\\
  c_{i,y}=\{0,0,0,0,0,1,1,-1,1,1,0,0,0,0,-1,-1,1,-1,-1\},\\
  c_{i,z}=\{0,0,1,1,1,0,1,1,0,0,0,-1,-1,-1,0,-1,-1,0,0\}.
\end{gathered}
\label{eq_cid3q19}
\end{equation}

The lattice weights $\omega_{i}$ of the $D3Q19$ scheme are given by
\begin{equation}
\begin{split}
  \omega_{i}=\left\{\frac{1}{3},\frac{1}{18},\frac{1}{36},\frac{1}{18},\frac{1}{36},\frac{1}{18},\frac{1}{36},\frac{1}{36},\frac{1}{36},\frac{1}{18},\frac{1}{36},\frac{1}{18},\frac{1}{18},\frac{1}{36},\right.\\
  \left.\frac{1}{18},\frac{1}{36},\frac{1}{36},\frac{1}{36},\frac{1}{36}\right\}.
\end{split}
\label{eq_wi}
\end{equation}

The equilibrium distribution function $f^{eq}_i$ is expressed as

\begin{equation}
    \begin{aligned}
        f_i^{eq} = \omega_i \bigg[ \rho 
            &+ \frac{\omega_i - \delta_{0i}}{\omega_i} \rho (\theta - 1) + \frac{\mathcal{H}_{i,\alpha}^{(1)}}{c_s^2} \rho u_\alpha + \frac{\mathcal{H}_{i,\alpha\beta}^{(2)}}{2c_s^4} 
            \rho u_\alpha u_\beta \\
            &+ \frac{\mathcal{H}_{i,\alpha\beta\gamma}^{(3r)}}{6c_s^6} 
            \rho u_\alpha u_\beta u_\gamma \bigg],
    \end{aligned}
\label{eq_feq}
\end{equation}
where $\theta= T/T_{ref}$ is the normalized temperature obtained from the resolution of the energy equation (\ref{eq_energy}). $T_{ref}$ is an arbitrary reference temperature used to adjust the CFL number \cite{wissocq2022restoring}. $c_s=\sqrt{RT_{ref}}=\Delta x/(\sqrt{3}\Delta t)$ is the lattice sound speed. $\mathcal{H}_i$ are discrete Hermite polynomials defined in Eqs. (\ref{eq_hi}) and (\ref{eq_hi3}).

The off-equilibrium distribution function $\bar{f}^{neq}_i$ is calculated by

\begin{equation}
\bar{f}_i^{neq}=\omega_i\left\{\frac{\mathcal{H}_{i,\alpha\beta}^{(2)}}{2c_s^4}\Pi_{\alpha\beta}^{{neq},(2)}+\frac{\mathcal{H}_{i,\alpha\beta\gamma}^{(3r)}}{6c_s^6}\Pi_{\alpha\beta\gamma}^{{neq},(3r)}\right\},
\label{eq_fneq}
\end{equation}
where $\Pi_{\alpha\beta}^{{neq},(2)}$ and $\Pi_{\alpha\beta\gamma}^{{neq},(3)}$ denote second- and third-order Hermite off-equilibrium coefficients, defined in Eqs. (\ref{eq_pi2}) and (\ref{eq_pi3}). These coefficients are evaluated using the hybrid recursive collision model proposed in \cite{jacob2018new}, which improves numerical stability.

The discrete Hermite polynomials $\mathcal{H}_i$ are calculated by
\begin{equation}
\begin{gathered}
    \mathcal{H}_i^{(0)} = 1,\\
    \mathcal{H}_{i,\alpha}^{(1)} = c_{i\alpha},\\
    \mathcal{H}_{i,\alpha\beta}^{(2)} = c_{i\alpha}c_{i\beta} - c_s^2\delta_{\alpha\beta},\\
    \mathcal{H}_{i,\alpha\beta\gamma}^{(3)} = c_{i\alpha}c_{i\beta}c_{i\gamma} - c_s^2(\delta_{\alpha\beta}c_{i\gamma} + \delta_{\beta\gamma}c_{i\alpha} + \delta_{\alpha\gamma}c_{i\beta}),
\end{gathered}
\label{eq_hi}
\end{equation}

\begin{equation}
\begin{gathered}
    \mathcal{H}_{i,1}^{(3r)} = \mathcal{H}_{i,xxy}^{(3)} + \mathcal{H}_{i,yzz}^{(3)},\\
    \mathcal{H}_{i,2}^{(3r)} = \mathcal{H}_{i,xzz}^{(3)} + \mathcal{H}_{i,xyy}^{(3)},\\
    \mathcal{H}_{i,3}^{(3r)} = \mathcal{H}_{i,yyz}^{(3)} + \mathcal{H}_{i,xxz}^{(3)},\\
    \mathcal{H}_{i,4}^{(3r)} = \mathcal{H}_{i,xxy}^{(3)} - \mathcal{H}_{i,yyz}^{(3)},\\
    \mathcal{H}_{i,5}^{(3r)} = \mathcal{H}_{i,xzz}^{(3)} - \mathcal{H}_{i,xyy}^{(3)},\\
    \mathcal{H}_{i,6}^{(3r)} = \mathcal{H}_{i,yyz}^{(3)} - \mathcal{H}_{i,xxz}^{(3)}.
\end{gathered}
\label{eq_hi3}
\end{equation}

The second-order Hermite off-equilibrium coefficient $ \mathrm{\Pi}_{\alpha\beta}^{{neq},(2)}$ is defined as
\begin{equation}
    \begin{aligned}
        \mathrm{\Pi}_{\alpha\beta}^{{neq},(2)}(x,t) = 
        &\ \tau \sum_{i=0}^{Q-1} 
        \left[
            c_{i\alpha} c_{i\beta} - \frac{\delta_{\alpha\beta}}{3} c_{i\gamma} c_{i\gamma}
        \right] \\
        &\left( 
            \bar{f}_i(x,t) - f^{eq}(x,t) + \frac{\Delta t}{2} F_i(x,t - \Delta t)
        \right) -\\
        & \Bigg[ 
            (1 - \tau)\rho c_s^2 \bar{\tau}
            \left(
                \frac{\partial u_\alpha}{\partial x_\beta} + \frac{\partial u_\beta}{\partial x_\alpha}
                - \frac{2\delta_{\alpha\beta}}{D} \frac{\partial u_\gamma}{\partial x_\gamma}
            \right)
        \Bigg],
    \end{aligned}
\label{eq_pi2}
\end{equation}
where $\tau$ is the weighting free parameter \cite{jacob2018new}, which is set to 0.98 in the present study. $D$ is the spatial dimension. The third-order Hermite off-equilibrium coefficient $\Pi_{\alpha\beta\gamma}^{{neq},(3)}$ is then calculated by
\begin{equation}
\Pi_{\alpha\beta\gamma}^{{neq},(3)}(x,t)=\left[u_\alpha\Pi_{\beta\gamma}^{{neq},(2)}+u_\beta\Pi_{\gamma\alpha}^{{neq},(2)}+u_\gamma\Pi_{\alpha\beta}^{{neq},(2)}\right](x,t).
\label{eq_pi3}
\end{equation}

The Hermite moment $a_{\alpha\beta}^{F,(2)}$ is defined as
\begin{equation}
\begin{aligned}
a_{\alpha\beta}^{F,(2)}=&\frac{2}{D}\delta_{\alpha\beta}\rho c_{s}^{2}\frac{\partial u_{\gamma}}{\partial x_{\gamma}}-\delta_{\alpha\beta}c_{s}^{2}\frac{\partial\rho(1-\theta)}{\partial t}+a_{\alpha\beta}^{C}+f_{u,\alpha}u_{\beta}+f_{u,\beta}u_{\alpha},
\end{aligned}
\label{eq_af2}
\end{equation}
where $a_{\alpha\beta}^{C}$ is the lattice-dependent component of the force term introduced in \cite{farag2021unified}.

The discrete operators $\delta_t$ and $\delta_\alpha$ are defined as
\begin{equation}
\begin{gathered}
\delta_t\Phi=\frac{\Phi(\boldsymbol{x},t+\Delta t)-\Phi(\boldsymbol{x},t)}{\Delta t},\\
\delta_\alpha\Phi=\frac{\Phi(\boldsymbol{x},t)-\Phi(\boldsymbol{x}-\boldsymbol{e}_\alpha\Delta x,t)}{\Delta x},
\end{gathered}
\label{eq_deltata}
\end{equation}
where $\boldsymbol{e}_\alpha$ is the unity vector in the direction $\alpha$.

The mass flux $F_{+\Delta \alpha/2}^{\rho}$ are defined by
\begin{equation}
\begin{aligned}
&F_{+\Delta x/2}^{\rho}(x,y,z)=\frac{\Delta x}{\Delta t}\Big[f_{1}^{col}(x,y,z)-f_{10}^{col}(x^{+},y,z)\\&+\frac12f_2^{col}(x,y,z^-)-\frac12f_{11}^{col}(x^+,y,z)+\frac12f_2^{col}(x,y,z)\\&-\frac12f_{11}^{col}(x^+,y,z^+)-\frac12f_4^{col}(x^+,y,z^-)+\frac12f_{13}^{col}(x,y,z)\\&-\frac12f_4^{col}(x^+,y,z)+\frac12f_{13}^{col}(x,y,z^+)+\frac12f_8^{col}(x,y^-,z)\\&-\frac12f_{17}^{col}(x^+,y,z)+\frac12f_8^{col}(x,y,z)-\frac12f_{17}^{col}(x^+,y^+,z)\\&-\frac12f_9^{col}(x^+,y^-,z)+\frac12f_{18}^{col}(x,y,z)-\frac12f_9^{col}(x^+,y,z)\\&+\frac12f_{18}^{col}(x,y^+,z)\Big],
\end{aligned}
\label{eq_fluxx}
\end{equation}

\begin{equation}
\begin{aligned}
&F_{+\Delta y/2}^{\rho}(x,y,z)=\frac{\Delta x}{\Delta t}\Big[f_{5}^{col}(x,y,z)-f_{14}^{col}(x,y^{+},z)\\&+\frac12f_8^{col}(x,y,z)-\frac12f_{17}^{col}(x^+,y^+,z)+\frac12f_8^{col}(x^-,y,z)\\&-\frac12f_{17}^{col}(x,y^+,z)+\frac12f_9^{col}(x,y,z)-\frac12f_{18}^{col}(x^-,y^+,z)\\&+\frac12f_9^{col}(x^+,y,z)-\frac12f_{18}^{col}(x,y^+,z)+\frac12f_6^{col}(x,y,z^-)\\&-\frac12f_{15}^{col}(x,y^+,z)+\frac12f_6^{col}(x,y,z)-\frac12f_{15}^{col}(x,y^+,z^+)\\&-\frac12f_7^{col}(x,y^+,z^-)+\frac12f_{16}^{col}(x,y,z)-\frac12f_7^{col}(x,y^+,z)\\&+\frac12f_{16}^{col}(x,y,z^+)\Big],
\end{aligned}
\label{eq_fluxy}
\end{equation}

\begin{equation}
\begin{aligned}
&F_{+\Delta z/2}^{\rho}(x,y,z)=\frac{\Delta x}{\Delta t}\biggl[f_{3}^{col}(x,y,z)-f_{12}^{col}(x,y,z^{+})\\&+\frac12f_2^{col}(x,y,z)-\frac12f_{11}^{col}(x^+,y,z^+)+\frac12f_2^{col}(x^-,y,z)\\&-\frac12f_{11}^{col}(x,y,z^+)+\frac12f_4^{col}(x,y,z)-\frac12f_{13}^{col}(x^-,y,z^+)\\&+\frac12f_4^{col}(x^+,y,z)-\frac12f_{13}^{col}(x,y,z^+)+\frac12f_6^{col}(x,y,z)\\&-\frac12f_{15}^{col}(x,y^+,z^+)+\frac12f_{6}^{col}(x,y^-,z)-\frac12f_{15}^{col}(x,y,z^+)\\&+\frac12f_7^{col}(x,y,z)-\frac12f_{16}^{col}(x,y^-,z^+)+\frac12f_7^{col}(x,y^+,z)\\&-\frac12f_{16}^{col}(x,y,z^+)\Big],
\end{aligned}
\label{eq_fluxz}
\end{equation}
where $x^\pm=x\pm\Delta x $ and $y^\pm=y\pm\Delta x$. The expressions of the momentum flux $F_{+\Delta \alpha/2}^{\rho u_{\alpha}}$ can then be obtained straightforwardly by replacing each $f^{col}_i$ in $F_{+\Delta \alpha/2}^{\rho}$ by $c_{i,\alpha}f^{col}_i$.

The linear function $\mathscr{F}_{+\Delta\boldsymbol{\alpha}/2}^*$ is given by
\begin{equation}
\mathscr{F}_{+\Delta\boldsymbol{\alpha}/2}^*(\Phi)=\begin{cases}\overline{\Phi}_{+\Delta\boldsymbol{\alpha}/2}(\boldsymbol{x},t) &\mathrm{if}\quad\tilde{\boldsymbol{u}}_{\alpha}\geq0,\\
\overline{\Phi}_{-\Delta\boldsymbol{\alpha}/2}(\boldsymbol{x}+\boldsymbol{e}_{\alpha}\Delta x,t)&\mathrm{else},\end{cases}
\label{eq_linear}
\end{equation}
where $\tilde{\boldsymbol{u}}_{\boldsymbol{\alpha}}$ is defined as $\tilde{\boldsymbol{u}}_{\alpha}=[\boldsymbol{u}_{\alpha}(\boldsymbol{x},t)+\boldsymbol{u}_{\alpha}(\boldsymbol{x}+\boldsymbol{e}_{\alpha}\Delta x,t)]/2$ to ensure the symmetry of the algorithm,
\begin{equation}
\begin{gathered}
\overline{\Phi}_{+\Delta\alpha/2}=\Phi_{+\Delta\alpha/2}+\frac{\boldsymbol{u}_\alpha}{2}\frac{\Delta t}{\Delta x}\left(\Phi_{-\Delta\alpha/2}-\Phi_{+\Delta\alpha/2}\right), \\
\overline{\Phi}_{-\Delta\alpha/2}=\Phi_{-\Delta\alpha/2}+\frac{\boldsymbol{u}_\alpha}{2}\frac{\Delta t}{\Delta x}\left(\Phi_{-\Delta\alpha/2}-\Phi_{+\Delta\alpha/2}\right), \\
\Phi_{+\Delta\alpha/2}=\Phi+\frac{\Delta_{\alpha}}{2},\quad\Phi_{-\Delta\alpha/2}=\Phi-\frac{\Delta_{\alpha}}{2},
\end{gathered}
\end{equation}
where $\Delta_{\alpha}$ approximates the slope of $\Phi$ in the $\alpha$-direction, and is given by
\begin{equation}
\begin{aligned}
\Delta_{\alpha}=&\frac12\Big\{(1+\boldsymbol{\eta}_\alpha)[\Phi(\boldsymbol{x},t)-\Phi(\boldsymbol{x}-\boldsymbol{e}_\alpha\Delta x,t)]\\&+(1-\boldsymbol{\eta}_\alpha)[\Phi(\boldsymbol{x}+\boldsymbol{e}_\alpha\Delta x,t)-\Phi(\boldsymbol{x},t)]\Big\},
\end{aligned}
\end{equation}
where $\boldsymbol{\eta}_{\alpha}$ is defined as $\boldsymbol{\eta}_{\alpha}=\frac{1}{3}\left[2\boldsymbol{u}_{\alpha}\frac{\Delta t}{\Delta x}-\mathrm{sign}(\boldsymbol{u}_{\alpha})\right]$ to obtain  a third-order accurate convection scheme in space and time \cite{wissocq2022restoring}.

\bibliographystyle{elsarticle} 
\bibliography{elsarticle}

\end{document}